\documentclass[twocolumn,trackchanges]{aastex7}

\usepackage{tabularx}
\usepackage{graphicx, amsmath, siunitx, multirow, caption, subcaption, makecell}
\usepackage{hyperref}
\usepackage{pgfplots}
\pgfplotsset{compat=1.18}

\usepackage[table,xcdraw]{xcolor}
\usepackage{tikz}
\usepackage{upgreek}

\begin{document}

\title{Impacts of Correlated Noise on Retrievals of Exo-Earth Atmospheres}

\author[orcid=0009-0006-1306-0928]{Nicole Wolff}
\affiliation{Department of Astronomy \& Astrophysics, University of California, Santa Cruz, Santa Cruz, CA 95060, USA}\email{newolff@ucsc.edu}

\author[0000-0003-1212-7538]{Bruce Macintosh}
\affiliation{Department of Astronomy \& Astrophysics, University of California, Santa Cruz, Santa Cruz, CA 95060, USA}
\affiliation{University of California Observatories, Santa Cruz, CA 95060, USA}
\email{}

\author[0000-0002-3196-414X]{Tyler D. Robinson}
\affiliation{Lunar \& Planetary Laboratory, University of Arizona, Tucson, AZ 85721, USA}
\email{}

\author[orcid=0000-0002-3199-2888]{Sarah Blunt}
\affiliation{Department of Astronomy \& Astrophysics, University of California, Santa Cruz, Santa Cruz, CA 95060, USA}
\email{}

\author[0000-0003-2233-4821]{Jean-Baptiste Ruffio}
\affiliation{Department of Astronomy \& Astrophysics,  University of California, San Diego, La Jolla, CA 92093, USA}
\email{}

\author[orcid=0009-0005-7704-5527]{Beck Dacus}
\affiliation{Department of Astronomy \& Astrophysics,  University of California, San Diego, La Jolla, CA 92093, USA}
\email{}

\author[0000-0001-7443-6550]{Alex Madurowicz}
\affiliation{Space Telescope Science Institute, Baltimore, MD 21218, USA}
\email{}

\author[0000-0002-3191-8151]{Marshall Perrin}
\affiliation{Space Telescope Science Institute, Baltimore, MD 21218, USA}
\email{}

\author[0000-0003-3818-408X]{Laurent Pueyo}
\affiliation{Space Telescope Science Institute, Baltimore, MD 21218, USA}
\email{}

\begin{abstract}
The Habitable Worlds Observatory (HWO) aims to accomplish high-contrast imaging and spectroscopy of true Earth analogs for the first time. However, high-contrast spectroscopy with HWO may be limited by residual speckles which persist after deformable mirror correction and post-processing, hindering atmospheric characterization. Previous studies of self-luminous giant planets showed that neglecting spectrally correlated errors due to speckles results in biased inferences of planetary parameters. Currently, HWO retrieval studies generate spectra without noise spectral correlations. We present a Gaussian Process model of correlated noise of known properties in simulated HWO exo-Earth spectra, and the integration of this noise source into the forward and inverse modeling tool \texttt{rfast}. We quantify the impact of correlated noise on inferred molecular abundances and planetary properties across the ultraviolet/visible/near-infrared bandpass, for varying spectral resolutions (R), signal-to-noise ratios (SNR), and noise correlation length-scales (L). We find that, at the fiducial UV/Vis/NIR R=7/140/70 and SNR=10, including spectrally correlated noise at L=200 nm and L=10 nm yields a 57\% and 161\% higher average uncertainty on log oxygen abundance, compared to uncorrelated noise. Consequently, it is critical for both instrument design and post-processing algorithms to minimize short length-scale chromaticity. Additionally, we find that a moderate resolution can constrain carbon dioxide (R$\geq 280$) and weakly detect  methane (R$\geq2800$) abundances, demonstrating the benefit of a moderate resolution near-infrared spectrograph. These findings can aid the interpretation of future HWO reflectance spectra and set requirements on optical quality, instrument stability, and observing configurations.

\end{abstract}

\keywords{\uat{Exoplanet astronomy}{486} --- \uat{Exoplanet atmospheres}{487} --- \uat{Direct imaging}{387}}

\section{Introduction} 

\subsection{The path to Earth-like exoplanets}
The field of exoplanet astronomy originated and exponentially expanded in the past quarter century, revealing populations of exoplanets unlike any in the Solar System. Yet Earth-mass exoplanets in the habitable zones of Sun-like stars, or ``Earth analogs,'' have remained largely undiscovered due to observational biases and fundamental technique limitations. For instance, the transit method has yielded an impressive quantity of discoveries of low-mass exoplanets at close separations from their host stars, especially with the Kepler (\citealt{Borucki+11}, \citealt{Dressing+13}) and TESS \citep{Ricker+15} missions. However, this technique is biased towards planets in the habitable zones of lower-mass stars, rather than G-type stars. Additionally, transit geometry becomes more improbable with increasing orbital separation, and requires a longer time baseline to follow-up a detection. Despite its limitations, the transit method will remain valuable in the search for Earth analogs, and ESA's PLATO mission is designed for this science case \citep{Rauer+25}. Even if Earth analogs are detected in a transiting geometry, there are limitations to characterizing their atmospheres via transmission spectroscopy. For one, the larger stellar diameter of a G-type star makes the planet's atmospheric signature relatively smaller and more difficult to disentangle from noise due to stellar activity \citep{Rackman+23}. In addition, transmission spectroscopy utilizes a slant observing  geometry \citep{Fortney+05} which requires starlight to pass through more of the atmosphere, and thus photons are attenuated at higher altitudes. Consequently, transmission spectroscopy is sensitive to upper atmospheres, and struggles to probe altitudes below \~15 km for Earth analogs primarily due to refraction effects \citep{Betremieux+13}.

It is of interest to characterize the near-surface atmospheres and surface features of Earth analogs in order to detect biosignatures and contextualize them with the planet's climate and habitability. The vertical distribution of gases in Earth's atmosphere is non-uniform; molecular water vapor is most abundant below the atmosphere's scale height of ~8.5 km \citep{Kindel+15}, and as a result, most water clouds form in the troposphere. Clouds can enhance the detectability of molecular gases, allowing for robust atmospheric characterization \citep{Kelkar+25}. In order to study the atmospheric constituents, cloud properties, and surface albedo of an exo-Earth, an essential observational technique will be direct reflected-light spectroscopy \citep{LUVOIR19}. In recent and coming years, much work has and will be focused in the field of direct imaging (or high-contrast imaging) and direct spectroscopy with the intention of both detecting and characterizing Earth analogs. The National Academy of Sciences 2020 Decadal Survey recommended a ``large infrared/optical/ultraviolet space telescope'' intended for the direct detection and spectral characterization of Earths around Sun-like stars \citep{Decadal2020}. This recommendation has led to the Habitable Worlds Observatory (HWO), a NASA mission under development for a launch in the late 2030s or early 2040s.

Throughout the mission concept evolution for HWO, atmospheric retrievals (Bayesian parameter inference from spectral data) have been and will be a necessary technique for exploring the science that the mission might achieve. For the initial HabEx and LUVOIR concepts (\citealt{Habex20},\citealt{LUVOIR19}), a notable early framework for retrievals over the visible bandpass (0.4-1.0 microns) is presented in \cite{Feng+18}. They conclude  that a resolving power, R= $\frac{\lambda}{\Delta \lambda}$, of 140 and a signal-to-noise ratio (SNR) per resolution element of 10 to 20 is necessary for ``meaningful'' constraints on oxygen and water abundance (i.e. a 1-sigma confidence interval less than an order of magnitude). Subsequent works have explored the prospects of constraining molecular abundances aside from oxygen and water. Ozone constraints likely require a low resolution of $\mathrm{R}=7$ in the UV \citep{Damiano+23}; detecting nitrous oxide on a Proterozoic Earth is possible at high enough abundances and at a wavelength cut-off of at least 1.4 microns \citep{Tokadjian+24}; the chemical disequilibrium of ozone and methane, a robust biosignature indicating biological forcing, can be detected on a Proterozoic Earth at an SNR of 50 \citep{Young+24}; and ruling out photochemical or geophysical false positives for biosignature detections likely requires a broad wavelength coverage and SNR of 20-40 \citep{krissansentotton+25}. Additional retrieval studies have simulated more realistic narrow-bandpass coronagraphic observations, and identify the optimal spectral regions for various molecules of interest \citep{Latouf+23}. To this point, \cite{Salvador+Robinson25} find that increasing the spectral coverage often yields greater benefits over increasing SNR. Importantly, the literature already highlights that retrievals are a necessary framework for these HWO trade studies because they result in estimates of abundance uncertainty, facilitating measurements of the detection significance of an observation (e.g. \citealt{krissansentotton+25}).

\subsection{Systematic noise in exoplanet imaging}
The HWO coronagraph and spectrograph for exo-Earth science is in the pre-conceptual design phase, motivating the trade study in this work. This instrument will leverage the lessons learned from existing high-contrast imagers and spectrographs, both ground-based (Gemini/GPI, \citealt{Macintosh+14}; VLT/SPHERE, \citealt{Beuzit+19}; Subaru/CHARIS, \citealt{Groff+17}) and space-based (JWST/NIRSpec, \citealt{Ruffio+24}). The past two decades of exoplanet science with these instruments and their predecessors have revealed the systematics and technical limitations that must be overcome to achieve a raw contrast of $10^{-10}$.

Assuming that HWO will have an ultra-low-noise detector \citep{LUVOIR19}, detector noise is not expected to be HWO's limiting noise source at spectral resolutions R $\lesssim$ 1,000 (\citealt{Wang+17}, \citealt{Ruffio+26_earths}). Rather, in this regime the limiting noise source will be primarily from exozodiacal light, with a secondary and non-negligible contribution to the noise budget being from stellar leakage, or speckle noise \citep{Ruffio+26_earths}. Speckles are diffraction-limited stellar point spread functions scattered across an image. For ground-based high-contrast imaging, speckles largely arise from wavefront error due to turbulence in Earth's atmosphere. Some residual speckles persist even after deformable mirror correction and post-processing. As diffraction effects, speckles are correlated in the spectral domain. Observations from ground-based IFUs indicate that the spectral covariance induced by speckles can be adequately described by a Gaussian  (\citealt{Greco+16}). This means that the variability in a spectrum induced by a speckle can also be described as a Gaussian (Appendix \ref{sec:psfmag}). Detrimentally, this means that residual speckles can create both false positives for planet detection and false spectral signatures for atmospheric characterization. 

Although space-based direct imaging can achieve greater sensitivity to faint planets by avoiding Earth's atmosphere, these observations will contain residual speckles as a source of spectrally correlated noise. For one, the telescope will be subject to thermal vibrations, inducing time-variable aberrations which will require active wavefront sensing and deformable mirror control (e.g. as is the case for JWST, \citealt{Acton+22}). Thus, wavefront reconstruction will be subject to similar time-variable imperfections to those of ground-based imaging. The optical stability requirement of the HWO coronagraph will be, in part, dependent on the tolerance of spectral inference and atmospheric retrievals to spectrally correlated noise.  

Additionally, optical polishing errors that are much larger than the wavelength of light will randomize coherent starlight, resulting in wavefront amplitude aberrations which remain uncorrected by deformable mirrors. This effect, known as the Talbot Effect, is wavelength-dependent, and leaves a residual term upon PSF subtraction with spectral differential imaging \citep{Marois+06}. We discuss the Talbot Effect in Section \ref{sec:correlated_noise}. Spectral correlations can also arise from systematics in the spectral extraction of IFU data \citep{Ruffio+26}. 

\subsection{Impacts of speckle noise on retrievals}

Retrievals for HWO have yet to explore the impacts of correlated noise. These considerations are likely to be important as earlier work in ground-based high contrast imaging have shown the importance of accounting for correlated noise. For example, previous spectroscopic observations of giant planets with the GPI and SPHERE integral field spectrographs have illustrated that the inclusion of an appropriate covariance matrix in atmospheric retrievals greatly impacts the resulting posterior distributions (\citealt{Greco+16}, \citealt{Nasedkin+23}). Specifically, failing to account for speckle noise both biases retrieved parameters and artificially reduces parameter uncertainty. \cite{Nasedkin+23} additionally find that this bias is significantly influenced by the PSF subtraction algorithm.

In order to compute the covariance matrix, it is common practice to assume a fixed functional form of the noise dependent on the wavelength spacing, F($\Delta\lambda$), and fit a spectrum of the noise to a covariance model such as a Gaussian \citep{Greco+16}. An alternative method is to compute an autocorrelation of the residuals between data and model in order to measure the covariance matrix, e.g. \cite{Madurowicz+25}. It is important to note that these methods are both satisfactory approximations, but neither allow for perfect measurement of the correlated noise.

To further correct for spectrally correlated noise, previous analyses have jointly modeled the spectrum and the correlated noise properties. This has been shown to recover a spectral correlation length for various applications, including low-resolution direct spectroscopy (\citealt{Wang+20}, \citealt{Xuan+22}), milliarcsecond astrometry (\citealt{Wang+16}), transmission spectroscopy (\citealt{Fortune+24}), and kernel phase interferometry with GRAVITY (\citealt{Thompson+25}).

\subsection{Outline}
The goal of this work is to estimate the scientific return of HWO for Earth analogs given different instrument designs (spectral resolution, bandpass) and observing configurations (signal-to-noise). This work builds upon previous studies by including the contribution from correlated speckle noise. We use a simple parametric model for the correlated noise because its specific properties will depend on the HWO spectrograph designs, which have not been finalized. By doing this, we aim to facilitate eventual end-to-end yield studies, allowing the science to drive the instrument design. 

In Section \ref{sec:methods}, we present a Gaussian Process parameterization of correlated noise in exo-Earth spectra observed by HWO, and the integration of this noise source into the forward modeling and retrieval tool \texttt{rfast}. Retrieval results spanning noise properties, spectral resolutions, and signal-to-noise ratios are presented in Section \ref{sec:results}. We discuss the implications of speckle noise on exo-Earth characterization and future steps in Section \ref{sec:discussion}, and conclude in Section \ref{sec:conclusions}.

\section{Methods}\label{sec:methods}

\subsection{Forward model}
Reflectance spectra are generated using the \texttt{rfast} package, a tool designed for rapid HWO trade studies consisting of a forward model and retrieval framework \citep{Robinson+Salvador23}. The \texttt{rfast} phase-independent module (i.e., its efficient ``diffuse'' mode), used here, simplifies the computation by performing one-dimensional radiative transfer assuming a plane-parallel atmosphere. The planet's albedo (A) is calculated by the plane-parallel solver assuming a Lambertian sphere. 
% Planetary albedo depends on surface albedo as the lower boundary condition, and, at each layer, the extinction optical depth ($\tau$) and single scattering albedo ($\Omega$).
At each layer in the atmosphere, reflectivity and transmittivity are computed by solving the radiative transfer equations. Then, the albedo ($\mathrm{A}_\mathrm{p}$) of the full column is recursively calculated from the preceding columns. 
% The lower boundary condition for column reflectivity is the albedo ($A_\mathrm{s}$). 
The planet-to-star flux ratio, $\frac{\mathrm{F}_\mathrm{p}}{\mathrm{F}_\mathrm{s}}$ (otherwise known as ``planet-to-star contrast"), depends on the stellar flux, the planetary flux, the planetary albedo, the planetary radius, and the semi-major axis (a): 

\begin{equation}
\frac{\mathrm{F}_\mathrm{p}}{\mathrm{F}_\mathrm{s}} = \mathrm{A}_\mathrm{}\left(\frac{\mathrm{R}_\mathrm{p}}{\mathrm{a}}\right)^2,
\end{equation}

The planetary atmosphere is simplified to be an isotherm of T=294 K. As an Earth analog, the planet has a radius, mass (equivalently surface gravity), and orbital separation equivalent to that of Earth. These retrievals are without clouds to reduce the forward model computation time by a factor of two; the implications of this choice are discussed in Section \ref{sec:discussion}. The Lambertian surface albedo is set to 0.2 to account for the average cloud coverage of an Earth analog \citep{Robinson26}. Opacities are from the HITRAN2020 molecular database \citep{Gordon+22} and are on a  1\,cm$^{-1}$ grid for the lower resolving power retrievals (this fixed resolution corresponds to $R\sim14,000$ at $700\,$nm). For higher resolving powers $(R\geq1000$), we use the higher-resolution 0.1\,cm$^{-1}$ grid (corresponding to $R\sim140,000$ at $700\,$nm). The planet's atmosphere is represented on a pressure grid ranging from 101,000 Pa at the surface to 1 Pa at the top. In this work, radiatively active gases are O$_2$, H$_2$O, CO$_2$, and in some of the cases presented here, CH$_4$, while N$_2$ is used as a radiatively inactive gas to fill the atmosphere to the appropriate pressure. The molecules that interact via collision-induced absorption (CIA) are N$_2$ and O$_2$. Rayleigh scattering is included. Therefore, the contributors to the total opacity are molecular absorbers, Rayleigh scattering, and CIA. These assumptions are similar to those in \cite{Feng+18} and \cite{krissansentotton+25}, except that for our preliminary study, we omit O$_3$ and clouds as opacity sources: O$_3$ because our focus is on design trades for the visible and infrared spectrographs rather than the likely second-generation UV coronagraph, and clouds because of the factor of two increase in computation time for radiative transfer. Clouds will be the subject of future work.

\subsection{Spectral observables of Earth}
\label{sec:observables}
\begin{figure}[h!]
 \begin{minipage}[c]{0.98\linewidth}
  \centering
  \begin{center}
  \includegraphics[height=14cm, angle=0]{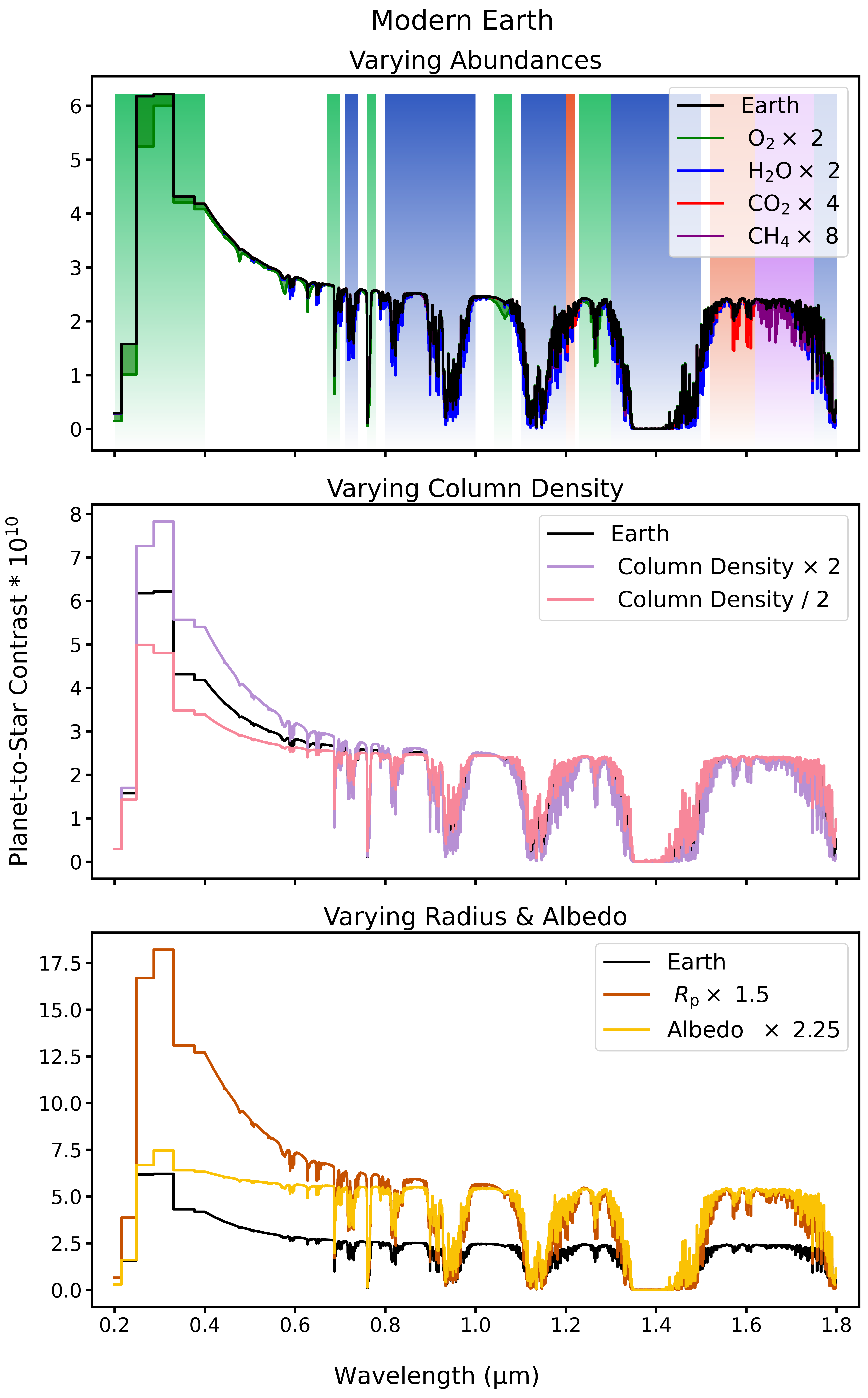}
  \vspace{-5pt}
  \caption{Theoretical spectra of Earth varying atmospheric and bulk properties, convolved to a resolution of 7 in the UV band and 1400 in the Visible/NIR bands. The fiducial spectrum of  Earth is shown in black. \textbf{Top:} Spectra given an increase in the molecular abundances of oxygen, water, and carbon dioxide. Wavelength regions sensitive to changing abundance are illustrated by the shaded bands. \textbf{Middle:} Spectra with varying column density. The column density is increased or decreased by scaling the planetary mass. \textbf{Bottom:} Spectra with increased planetary radius or surface albedo.}
  \label{f:exoearths}
  \vspace{-6pt}
  \end{center}
  \end{minipage}
\end{figure}

Figure \ref{f:exoearths} demonstrates the observables of bulk and atmospheric properties on Earth's spectrum. For non-saturated lines, increased abundances are directly proportional to spectral line depth. To illustrate the effects of bulk properties (radius, surface albedo, surface pressure) on the planet's spectrum, it helps to identify the degenerate parameters in the atmospheric model. 

One observable is spectral line (or band) depth and shape. To observe the effect of molecular abundances on spectral absorption features, we generate spectra with increased abundances of O$_2$, H$_2$O, CO$_2$, and CH$_4$, while scaling down the filler gas, N$_2$, to maintain a fixed column density. At Modern Earth abundances, the most observable and abundant molecules are O$_2$ and H$_2$O (see Table \ref{t:parameter_inputs} for their abundances), for which a factor of 2 increase in abundance yields an observable increase in band depth (Figure \ref{f:exoearths}, top). Because the fractional abundances of carbon dioxide and methane are lower, they must be scaled by more significant factors in order to observe a band depth change. Figure \ref{f:exoearths} (middle) illustrates the degeneracy between the surface pressure and these molecular abundances, because a spectral line can be deepened by either an increased fractional abundance or by increasing overall column density, which can be tuned by changing the planet's surface pressure. This degeneracy can be resolved by observing a spectral point from 300-900 nm, as the Rayleigh scattering slope is sensitive to the atmospheric column density. 

Another canonical pair of degenerate parameters is the radius and surface albedo: a brighter flux could either indicate a larger planet or a more reflective surface. As seen in Figure \ref{f:exoearths} (bottom), the radius-albedo degeneracy is broken at shorter wavelengths, as the surface albedo only influences spectral regions sensitive to surface features. This excludes the Rayleigh scattering region from being affected by the surface albedo \citep[][ Figure 3.1-3]{Habex20}. As a result, photometry or spectroscopy in the wavelength range from 300-900 nm is crucial to break the radius-albedo degeneracy. Note that the radius and albedo scaling factors in Figure \ref{f:exoearths}b are arbitrary and are simply chosen to produce an equivalent spectrum past 900 nm.

\begin{figure}[h!]
 \begin{minipage}[c]{1.0\linewidth}
  \centering
  \begin{center}
  \includegraphics[height=9.2cm, angle=0]{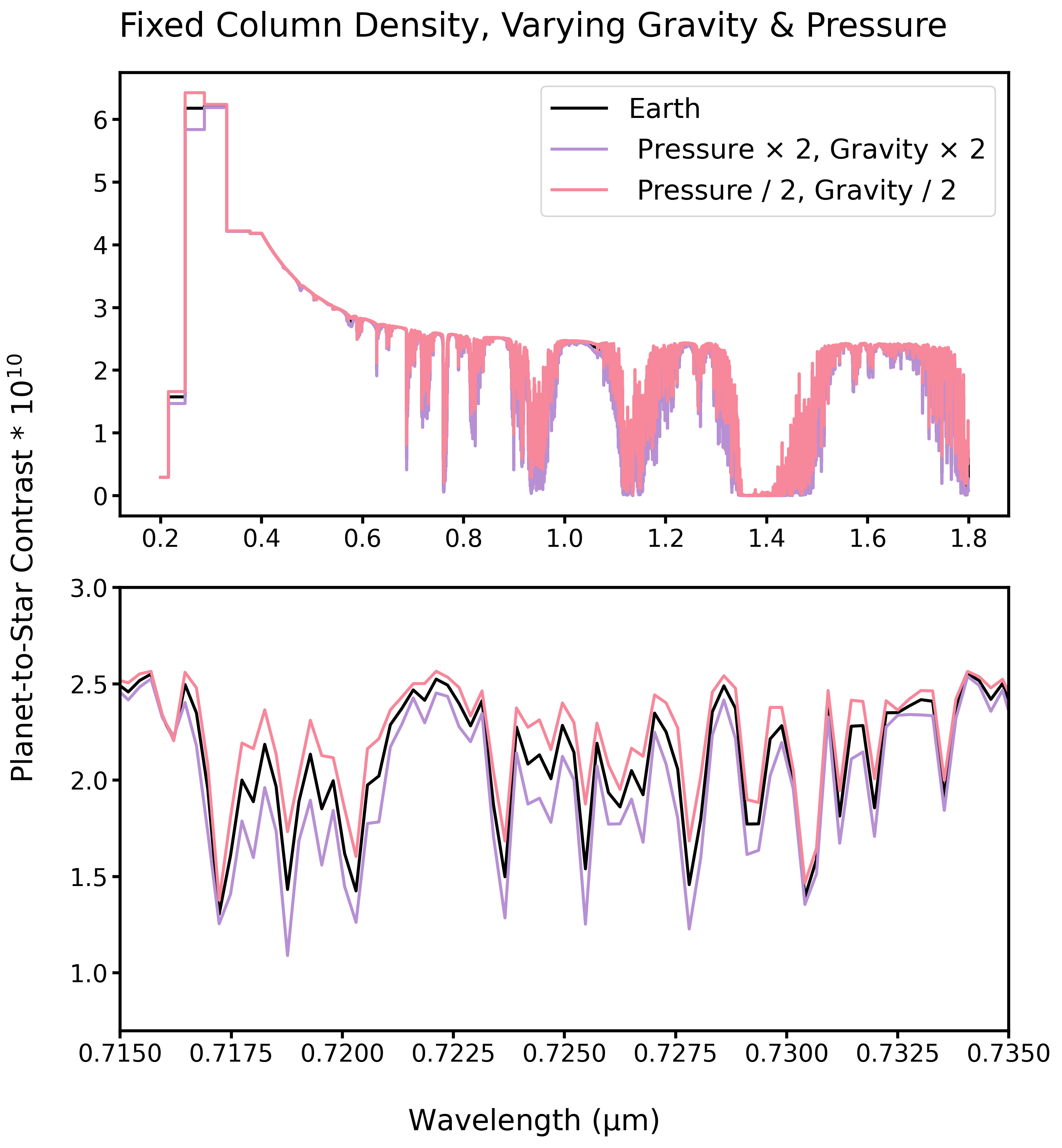}
  \vspace{-5pt}
  \caption{Theoretical spectra of Earth, convolved to a resolution of 7 in the UV band and 2800 across the Visible/NIR bands. \textbf{Top:} The purple spectrum illustrates a high surface pressure while pink illustrates a low pressure, both for the same column density. \textbf{Bottom:} Zoomed into show various H$_2$O doublets around 720 nm.}
  \label{f:pressure_broadening}
  \vspace{-6pt}
  \end{center}
  \end{minipage}
\end{figure}

Collisional/pressure broadening is evident in Figure \ref{f:pressure_broadening}, as spectral lines become broader due to the increased collisions in a higher pressure gas. Pressure broadening helps to partially break the degeneracy between surface pressure and planetary mass (or equivalently, surface gravity) via their relation to the column number density \citep[where additional information about atmospheric mean molar mass is required to fully break the degeneracy;][]{Damiano+25}. Doubling the pressure and doubling the surface gravity results in the same column density, but increasing the pressure uniquely alters the line shapes, resulting in deeper and broader bands.

\subsection{Correlated noise model}
Speckles are diffraction effects, and as such they magnify with wavelength, meaning that they move radially outward and increase in size. This phenomenon is referred to as PSF magnification and is detailed in Appendix ~\ref{sec:psfmag}. This static behavior of speckles is the simplest and the easiest to model. In reality, correlated noise arises from a multitude of sources, including fringing effects \citep{Lardiere+23}, optical polishing defects \citep{Krist+23}, telescope rotation for angular differential imaging, and IFU post-processing \citep{Ruffio+26_earths}. Eventually, noise error budgets for the HWO coronagraph will ideally include all the correlated noise sources and their spectral and temporal properties. 

\begin{figure*}[th!]
 \begin{minipage}[c]{1.0\linewidth}
  \centering
  \begin{center}
  \includegraphics[height=7cm, angle=0]{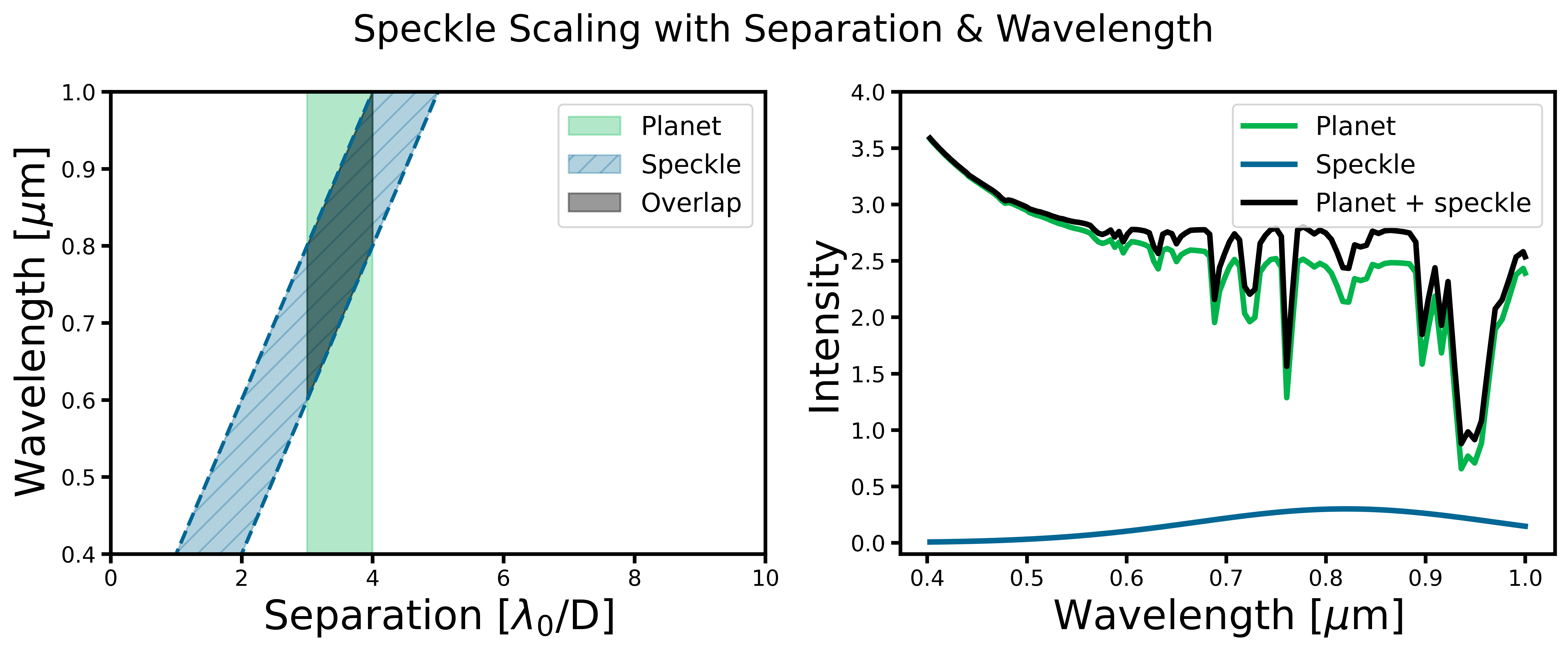}
  \vspace{-5pt}
  \caption{\textbf{Left:} Planets remain at fixed separations across multi-wavelength images, while speckles scale radially outward with wavelength. \textbf{Right:} In a spectrum, a speckle's radial scaling manifests as a Gaussian peaking at some wavelength, with some characteristic spectral length scale. This speckle contribution modifies the planetary flux across that length scale (black).}
  \label{f:speckle}
  \vspace{6pt}
  \end{center}
  \end{minipage}
\end{figure*}

In the absence of an existing instrument design, we choose to use a Gaussian model to approximate the effects of speckles in an extracted planet spectrum. In this model,  an individual speckle peaks in flux at the specific wavelength where it overlaps with the planet's point spread function (Figure \ref{f:speckle}).

To generate noisy spectra with noise in \texttt{rfast}, we first simulate an observation by convolving a high-resolution model with a Gaussian to degrade the spectral resolution. The noise contains a spectrally uncorrelated and a spectrally correlated component, which are combined in quadrature. The white noise, or the uncorrelated noise, is first added in quadrature to each data point. This is parameterized by an uncorrelated S/N ($\mathrm{SNR_{u}}$). This is calculated as the signal at $\lambda_0$ (550 nm for full bandpass retrievals, 680 nm for narrow 20\% bandpass retrievals) divided by the noise standard deviation:

\begin{equation}
\mathrm{SNR}_\mathrm{u} = \frac{\mathrm{F}_\mathrm{p}/\mathrm{F}_\mathrm{s}(\lambda_0)}{\alpha_\mathrm{u}}
\end{equation}

To generate correlated noise due to speckles, we use a Gaussian Process (GP) model \citep{Rasmussen+06}. Based off insight from previous work, e.g. a scientific yield study for Roman CGI which sampled different correlated noise parameterizations (\citealt{Lupu+16}), we use a squared exponential kernel, parameterized by amplitude ($\alpha_\mathrm{c}$) and length scale ($\sigma_\mathrm{c}$): 

\begin{equation} \label{e:cov_matrix}
 \text{K} (\lambda_i,\lambda_j) = \alpha_\mathrm{c}^2 \ \text{exp} \left[- \frac{(\lambda_i - \lambda_j)^2}{2\sigma_\text{c}^2}\right]
\end{equation}

\begin{figure}[h!]
 \begin{minipage}[c]{1.0\linewidth}
  \centering
  \begin{center}
  \includegraphics[height=10cm, angle=0]{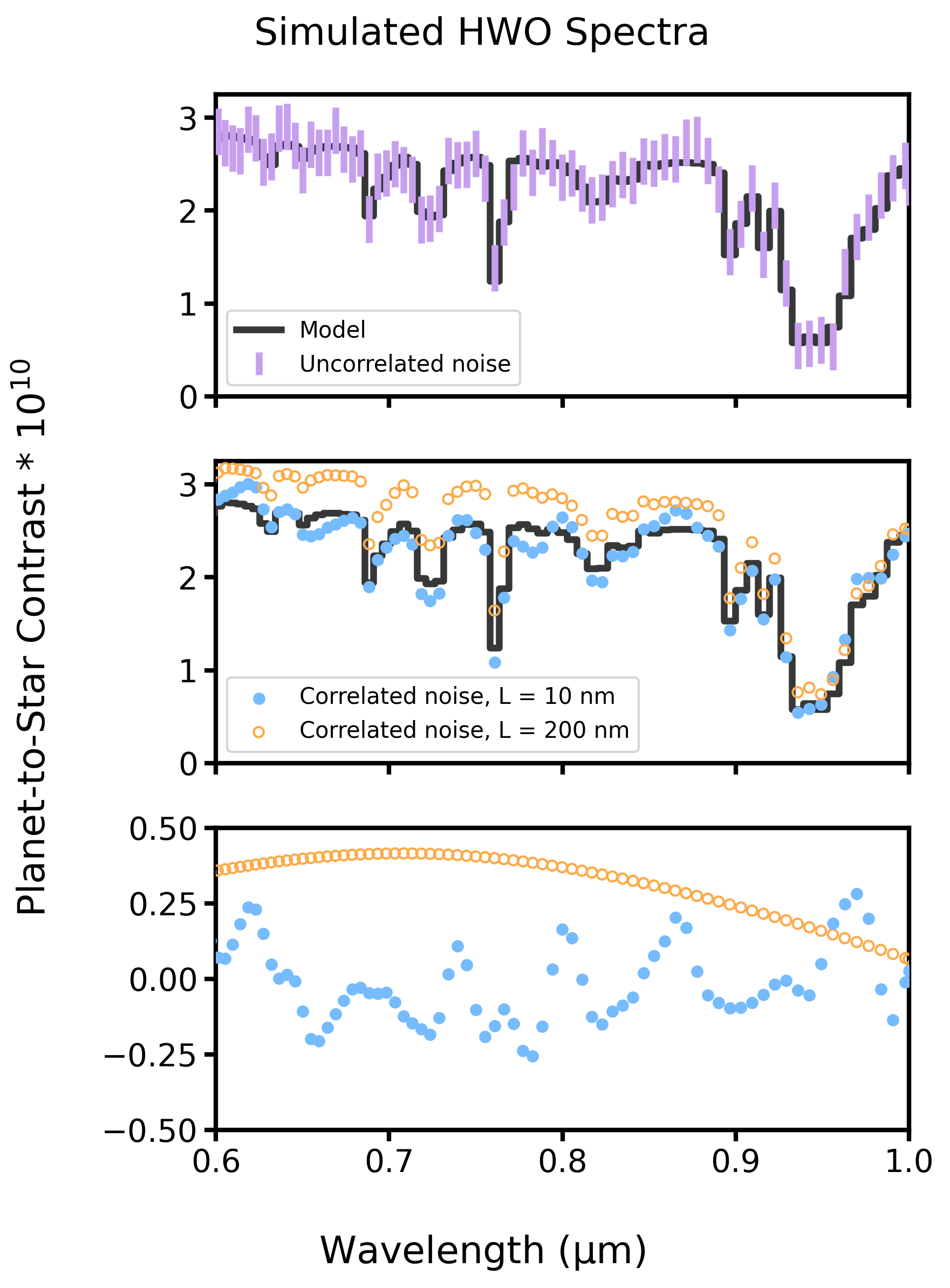}
  \vspace{-5pt}
  \caption{\textbf{Top:} A theoretical spectrum of Earth is plotted in black, across a region of the visible band from  0.6-1 $\upmu$m at R=140. Gaussian uncorrelated noise is added to the spectrum, with error bars shown in purple. \textbf{Middle:} Theoretical spectrum of Earth (black), with added correlated noise at a short length scale (blue, filled circles) and long length scale (orange, open circles). \textbf{Bottom:} Spectra of the correlated noise added in the middle panel.}
  \label{f:ex_corrnoise}
  \vspace{-6pt}
  \end{center}
  \end{minipage}
\end{figure}

Gaussian Processes enable sampling multiple random realizations of the speckle noise. Figure \ref{f:ex_corrnoise} exemplifies the effects of correlated noise on a spectrum of Earth from 0.6-1 $\upmu$m. In this figure, the longer 200 nm length scale scenario broadly affects the continuum, while 10 nm correlated noise either produces random scatter or modifies the depth of absorption features (e.g., the oxygen line at 0.68 $\upmu$m, or the water line at 0.72 $\upmu$m).

% \begin{figure*}[h!]
%  \begin{minipage}[c]{1.0\linewidth}
%   \centering
%   \begin{center}
%   \includegraphics[height=9cm, angle=0]{const_seed_noise_spectra.png}
%   \vspace{-5pt}
%   \caption{}
%   \label{f:noise_spectra_fixed_seed}
%   \vspace{-6pt}
%   \end{center}
%   \end{minipage}
% \end{figure*}

The amplitude of the correlated noise is parameterized by a correlated S/N ($\mathrm{SNR}_\mathrm{c}$). This S/N is calculated by the signal at  $\lambda_0$ (550 nm for full bandpass retrievals, 680 nm for narrow 20\% bandpass retrievals) divided by the noise standard deviation: 

\begin{equation}
\mathrm{SNR}_\mathrm{c} = \frac{\mathrm{F}_\mathrm{p}/\mathrm{F}_\mathrm{s}(\lambda_0)}{\alpha_\mathrm{c}}
\end{equation}

The total S/N, and thus the error bar on an individual spectral point, is thus given by the quadrature sum of uncorrelated and correlated S/N: 

\begin{equation}
\left(\frac{1}{\mathrm{SNR}}\right)^2 = \left(\frac{1}{\mathrm{SNR}_\mathrm{u}}\right)^2 + \left(\frac{1}{\mathrm{SNR}_\mathrm{c}}\right)^2
\label{e:snr_sum}
\end{equation}

The standard assumption in this work is $\mathrm{SNR}_\mathrm{c} = 10$ to probe correlated noise that is 10\% of the continuum. This SNR maps directly to the PSF stability in contrast units, which can connect to telescope vibration requirements. \cite{Ruffio+26_earths} translate this correlated SNR to a post-processing gain, i.e. the ability to subtract residual correlated starlight in post-processing. They perform a complementary investigation to this paper, developing a simulation toolkit to investigate the SNR for molecular detection as a function of spectral resolution, assuming realistic noise sources which include correlated speckle noise. They explore gains of [0.1, 0.01, 0.001]. Assuming the instrument achieves equivalent raw contrast to the planet-to-star flux ratio, these gains translate to $\mathrm{SNR}_\mathrm{c}$ = [10, 100, 1,000]. The assumption in this work that we adopt is the most pessimistic of \cite{Ruffio+26_earths}: a speckle contrast stability of $10^{-11}$ or a gain of 0.1, assuming $10^{-10}$ raw contrast for an Earth-Sun analog.

In one section of this paper (see Table \ref{t:all_retrievals}, ``Narrow BP, varying R''), $\mathrm{SNR}_\mathrm{c}$ = [2, 5, 10] are explored to probe speckle stability at the level of $\num{5e-11}$, $\num{2e-11}$, and $\num{1e-11}$, respectively. These are more pessimistic assumptions for the speckle stability, in order to understand scientific yield in a worst-case scenario for optical stability.

Once correlated noise has been added to the spectra with Gaussian Processes, one can also provide the MCMC with full information of the noise properties in the form of a covariance matrix. To do this, the likelihood function is updated to include a covariance matrix \citep{Czekala+15}, as in Equation \ref{e:log_likelihood}. In this case, we fix the Gaussian Process hyperparameters that describe the covariance matrix, rather than letting them be free parameters in the retrieval. This assumes that an observer has measured the S/N and length scale of the correlated noise, and wishes to account for the resulting uncertainty properly in their parameter estimation. \cite{Ruffio+24}, Appendix E exemplifies the autocorrelation technique for empirically measuring a covariance matrix for JWST/NIRSpec. 

\begin{figure*}[th!]
 \begin{minipage}[c]{1.0\linewidth}
  \centering
  \begin{center}
  \includegraphics[height=6.5cm, angle=0]{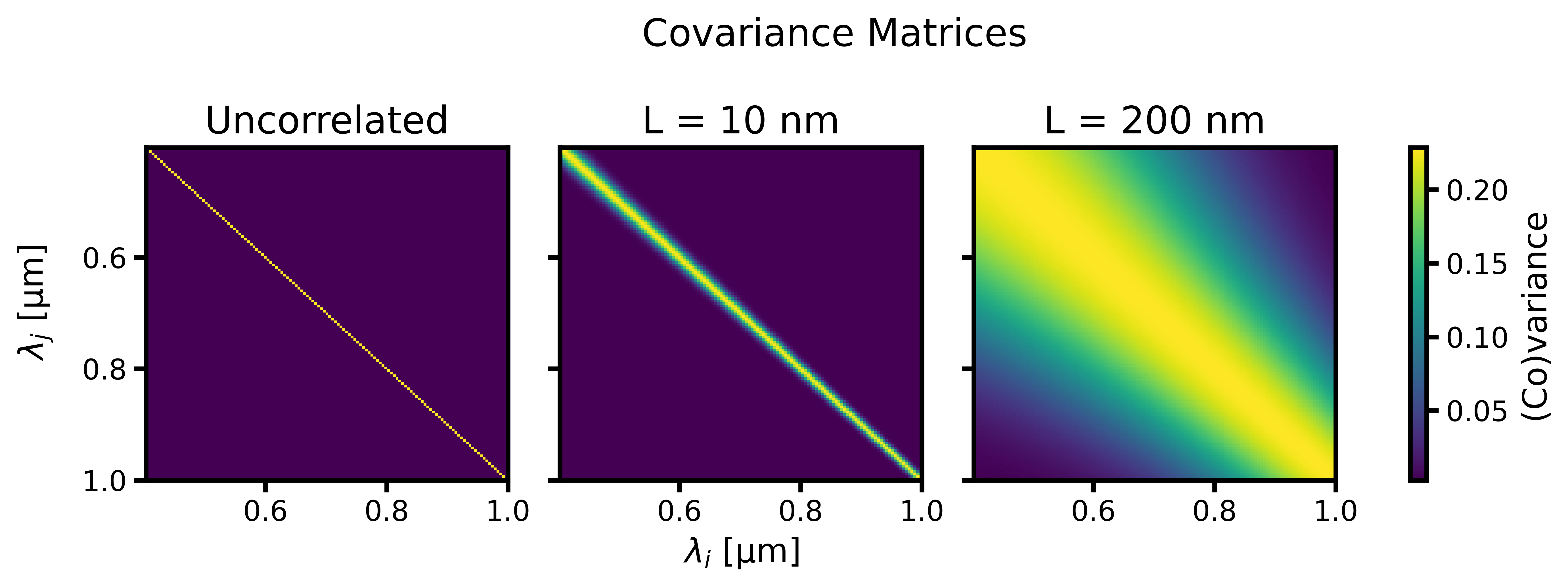}
  \vspace{-5pt}
  \caption{Example visible-band covariance matrices for three noise cases. \textbf{Left:} Covariance matrix for uncorrelated noise. The matrix is entirely diagonal. \textbf{Middle:} Covariance matrix for correlated noise with a 10 nm length scale. \textbf{Right:} Covariance matrix for correlated noise with a 200 nm length scale.}
  \label{f:cov_matrices}
  \vspace{-6pt}
  \end{center}
  \end{minipage}
\end{figure*}

Figure \ref{f:cov_matrices} illustrates an example of visible-band covariance matrices for a purely uncorrelated case and for two different length scales. The covariance changes with wavelength; for any two spectral points which are N resolution elements away, their covariance is larger at shorter wavelengths, as seen by the wider gradient in covariance. This occurs because we fix the spectral resolution across the band-pass, meaning that two adjacent points will have a shorter $\Delta\lambda$ for shorter $\lambda$, and by Equation \ref{e:cov_matrix}, this results in a stronger covariance. Fixing the spectral resolution across the band-pass is done in order to match previous retrieval assumptions e.g. \cite{Feng+18} and \cite{krissansentotton+25}, although in practice a spectrograph's resolving power will slightly vary with wavelength. Finally, in order to simulate covariances across the full band-pass, we construct individual covariance matrices for the UV, visible, and NIR bands and stitch them together, such that the covariance is zero between spectral points in different arms of the instrument (e.g. the visible and NIR IFUs are not correlated).

\subsection{Retrievals}
Retrievals are run in \texttt{rfast} using MCMC sampling from \texttt{emcee} \citep{ForemanMackey+13} using an affine-invariant ensemble sampler \citep{Goodman+Weare10}.  Each retrieval takes 100,000 steps and uses 50 walkers. To construct the final posterior, we apply thinning at an interval of 2 and a 95\% burn-in. The retrieval hyper-parameters are chosen by visual tests for convergence, and we ensure that the walkers flatten out in the parameter space in the final 5,000 steps.

For an Earth analog, the true bulk and atmospheric properties input to this model and subsequently retrieved are shown in Table \ref{t:parameter_inputs}. All abundances are quantified as volume mixing ratios (VMRs), and all priors are log-uniform.

\begin{deluxetable*}{c|c|c|c}
\centering
% \begin{tabular}{c|c|c|c}
\tablehead{\textbf{Parameter} & \textbf{Definition} & \textbf{Input} & \textbf{Prior}}
\startdata
\hline
$\mathrm{f}_{\mathrm{O}_2}$ & Oxygen VMR  & 0.21 & [$10^{-10}$, 1]  \\
\hline
$\mathrm{f}_{\mathrm{H}_2\mathrm{O}}$ & Water VMR  & \num{3e-3} & [$10^{-10}$, 1]    \\
\hline
$\mathrm{f}_{\mathrm{CO}_2}$ & Carbon-dioxide VMR  & \num{4e-4} & [$10^{-10}$, 1]   \\
\hline
$\mathrm{f}_{\mathrm{CH}_4}$ & Methane VMR  & \num{2e-6} & [$10^{-10}$, 1]   \\
\hline
$\mathrm{M}_\mathrm{p}$ & Mass [$\mathrm{M}_\mathrm{E}$]  & 1 & [0.1, 100]   \\
\hline
$\mathrm{P}_0$ & Surface pressure [Pa]  & \num{1.01e5} & [\num{1e0}, $10^8$]  \\
\hline
$\mathrm{R}_\mathrm{p}$ & Radius [$\mathrm{R}_\mathrm{E}$]  & 1 & [0.1, 10]  \\
\hline
$\mathrm{A}_\mathrm{s}$ & Surface albedo  & 0.2 & [0.01, \num{1e0}]   \\
\hline
% \end{tabular}
\enddata
\caption{Retrieved parameters in this study, along with their true values and prior distributions. ``VMR'' abbreviates volume mixing ratio, a measure of the fractional molecular abundance. All priors are log uniform.}
\label{t:parameter_inputs}
\end{deluxetable*}

Atmospheric retrieval studies most often perform maximum likelihood estimation using a Gaussian likelihood function assuming uncorrelated noise of the form:

\begin{equation}
       \text{ln} \space \mathcal{L} \space \propto -0.5 \left[ \frac{(\text{y} - \text{F(x)} )^{2}} { \text{$\sigma$}^2} \right] 
\end{equation}

F(x) represents the spectral model, y represents the data, and $\sigma^2$ is the uncorrelated noise. Accounting for the full correlation matrix of the noise, the Gaussian likelihood goes as: 

\begin{equation}
       \text{ln} \space \mathcal{L} \space \propto -0.5 [(\text{y} - \text{F(x)} )^{\text{T} } \space \text{C}^{-1} \space (\text{y} - \text{F(x)} ) ] 
\label{e:log_likelihood}
\end{equation}

This latter equation is adopted for the likelihood in our study.

\section{Retrieval Results} \label{sec:results}

Table \ref{t:all_retrievals} provides a description of all retrieval cases in this study. We explore retrieval results across two regimes.

\begin{deluxetable*}{c|c|c|c|c|c|c|c|c}
\centering
\tablehead{\textbf{Case} & \textbf{R} & \textbf{SNR}& \textbf{SNR}& \textbf{Length scale} & \textbf{Bandpass} & \textbf{Noise} & \textbf{Noise} & \textbf{Matrix} \\
 & & \textbf{uncorr.} & \textbf{corr.} & \textbf{[nm]} & & \textbf{realizations} & \textbf{added?} & \textbf{included?}}
\startdata
\hline
Matrix & 140 & 20 & 10 & 10 & Full & 1 & Y & N \\
inclusion & 140 & 20 & 10 & 10 & Full & 1 & Y & Y \\
\hline
Varying L & 140 & 20$\sqrt{2}$ & 10 & \makecell{[10,20,40,\\80,100,200]} & Full & 1 & N & Y \\
\hline
\multirow{3}{*}{Bias} & 140 & 10 & N/A & N/A & Full & 20 & Y & N/A \\
& 140 & 20$\sqrt{2}$ & 10 & 10  & Full & 20 & Y & Y \\
& 140 & 20$\sqrt{2}$ & 10 & 200 & Full & 20 & Y & Y \\
\hline
Narrow BP, & \makecell{[140,280,\\1400,2800]} & \makecell{Various,\\ R-\\dependent} & [2,5,10] & [50,200] & 20\% Vis & 1 & Y & Y \\
varying R & & & & & & & & \\
\hline
Full BP, &\makecell{[140,280,\\1400,2800]} & 20 & 10 & 10 & Full & 1 & N & Y \\
varying R & & & & & & & & \\
\hline
\enddata
\caption{The parameters of all retrieval cases in this work.}
\label{t:all_retrievals}
\end{deluxetable*}

The first regime is the low-spectral resolution standard assumption derived from the LUVOIR \& HabEx studies: R=7 (0.2-0.4 $\upmu$m), R=140 (0.4-1.0 $\upmu$m), R=70 (1.0-1.8 $\upmu$m). These nominal resolving powers were chosen in order to resolve ozone absorption in the UV, oxygen absorption in the visible, and water absorption in the NIR. 

For this baseline spectral resolution, we first aim to validate the correlated noise model. To do this, we run retrievals both omitting and including a covariance matrix, to ensure that the matrix inclusion case yields broader and less biased posteriors. Then, we run retrievals on spectra with correlated noise spanning various length scales, $\sigma_c \in$ [10, 20, 40, 80, 100, 200] nm. Given our noise model, we expect the covariance, and thus the parameter uncertainty, to scale inversely with length scale, given the $\sigma_c$ dependence in Equation \ref{e:cov_matrix}.

\begin{figure}[h!]
 \begin{minipage}[c]{1.0\linewidth}
  \centering
  \begin{center}
  \includegraphics[height=9cm]{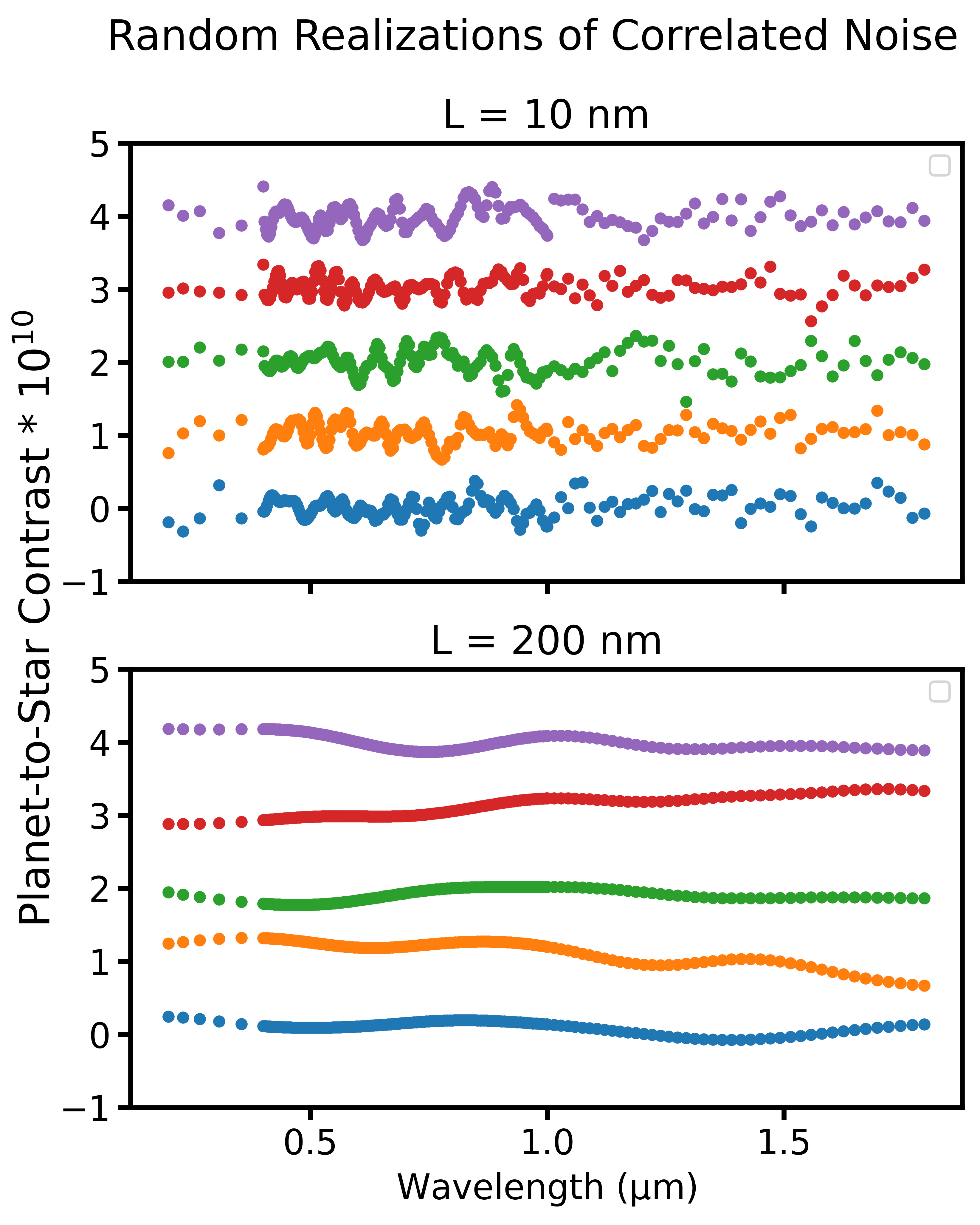}
  \vspace{-5pt}
  \caption{\textbf{Top:} 5 random draws of correlated noise with length scale of 10 nm. \textbf{Bottom:} 5 random draws of correlated noise with length scale of 200 nm.}
  \label{f:noise_spectra_20_instances}
  \vspace{-6pt}
  \end{center}
  \end{minipage}
\end{figure}

Next, we aim to investigate the potential biases that arise from inopportune realizations of the correlated noise, and whether there is a systematic bias in parameter inference regardless of the functional form of the correlated noise. We generate a statistical sample of 20 random realizations of the correlated noise for both a 10 nm and 200 nm length scale, and run retrievals. A subset of 5 of these noise spectra for each length scale are shown in Figure \ref{f:noise_spectra_20_instances}.

The second regime explored here is at varying spectral resolution and signal-to-noise, across a narrow 0.65-0.8 micron bandpass. First, we run retrievals across a narrow 20\% band-pass in the visible of 650-800 nm, encapsulating oxygen and water absorption lines. These retrievals effectively simulate a single coronagraphic channel with an optimistic bandpass, as in \cite{Latouf+23}. This is done for length scales of L=50 and L=200 nm. A 200 nm length scale is motivated mathematically by PSF magnification for an Earth-Sun system at a distance of 5 pc (Appendix \ref{sec:psfmag}). Shorter length scales can arise from PSF chromaticity, and thus we include a length scale of 50 nm, which matches the spectral width of the water line present in this bandpass and at redder wavelengths (Figure \ref{f:exoearths}). Values for ($\mathrm{SNR}_\mathrm{uncorr}$) are chosen to probe short, medium, and long exposure times. These SNRs scale down with spectral resolution assuming a photon-noise limited regime. This ensures that results at varying spectral resolutions can be directly compared, assuming an equivalent exposure time. For instance, in a given exposure time, a high-resolution (HR) spectrum will have a proportionally lower SNR per resolution element than a low-resolution (LR) spectrum, as the light is spread out over more pixels, as in Equation \ref{e:6}:

\begin{equation}\label{e:6}
\mathrm{SNR_{HR}} = \mathrm{SNR_{LR}}\sqrt{\frac{\mathrm{R_{low}}}{\mathrm{R_{high}}}}
\end{equation}

A photon-noise scaling is valid for the LUVOIR detector assumption up to R $\approx$ 1,000 \citep{Ruffio+26_earths}, after which detector noise likely will dominate. In this work, we adopt more optimistic assumptions of the detector noise, assuming a great improvement in detector technology, e.g. avalanche photodiode arrays (\citealt{Huber+24}). This allows us to isolate the effects of correlated noise without adopting a specific assumption of detector noise. The uncorrelated noise SNR per bin at each spectral resolution, in the photo noise-limited regime, is shown in Table \ref{t:r_snrs}.

\begin{deluxetable*}{c|c|c|c}
\centering
\tablehead{\textbf{R} & \textbf{Low SNR} & \textbf{Medium SNR} & \textbf{High SNR}}
\startdata
\hline
140 & 10  & 20 & 40  \\
\hline
280 & 7.07 & 14.14 & 28.28 \\
\hline
1400 & 3.16 & 6.32 & 12.65 \\ 
\hline
2800 & 2.24 & 4.47 & 8.94 \\
\hline
\enddata
\caption{Uncorrelated noise SNR per resolution element ($\mathrm{SNR}_\mathrm{u}$) at each spectral resolution, following Equation \ref{e:6}. These SNRs are the ``Equivalent Uncorrelated SNR'' reported in Figure \ref{f:narrow_bandpass_1}.}
\label{t:r_snrs}
\end{deluxetable*}

For these retrievals, we keep constant the functional form of the noise spectrum (i.e. choosing one noise realization from Figure \ref{f:noise_spectra_20_instances}), scaling the amplitude to vary the S/N and interpolating the spectral points to vary the spectral resolution. We do this for both L=200 nm and L=50 nm. In both resulting noise realizations, the spectrum is positive, resulting in an additive contribution to the flux. Given that the functional form of the noise is fixed, we do not generate random noise realizations, and can more efficiently explore the effects of correlated noise across the (R, SNR) parameter space.

The final retrieval cases we run are across the full band-pass at various spectral resolutions, and for one slice of the SNR parameter space: $(\mathrm{SNR}_\mathrm{u}, \mathrm{SNR}_\mathrm{c}) = (20,10)$. These are intended to explore the detectability of features in the near-infrared, CO$_2$ and CH$_4$, as a function of resolution, in the presence of realistic correlated noise.

\subsection{Inclusion of covariance matrix}
\begin{figure*}[t!]
 \begin{minipage}[c]{1.0\linewidth}
  \centering
  \begin{center}
  \includegraphics[width=\linewidth, angle=0]{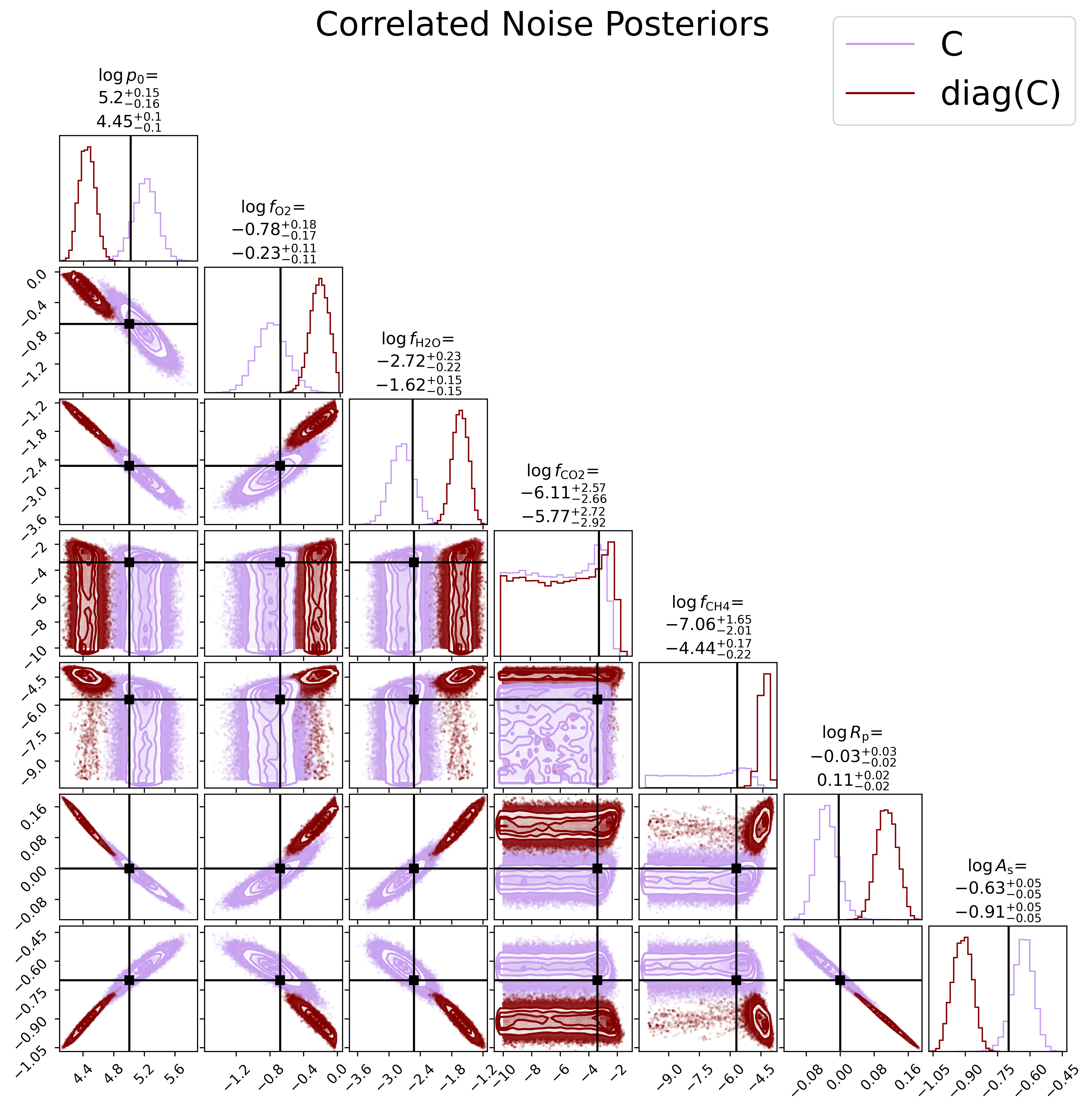}
  \vspace{-5pt}
  \caption{Corner plot of 1-D and 2-D posterior distributions. Both cases shown here include correlated noise in the spectrum. Retrieval results that excludes the covariance matrix (red) are plotted over results that include the matrix (purple). 1$\sigma$ confidence intervals are above each parameter for the full matrix case (top) and the no matrix case (bottom). The black vertical line indicates ground truth.}
  \label{f:no_cov_matrix}
  \vspace{-6pt}
  \end{center}
  \end{minipage}
\end{figure*}

In Figure \ref{f:no_cov_matrix}, we plot both one- and two-dimensional marginalized posteriors of each retrieved parameter for our first retrieval case. In the ``no matrix'' scenario, we add both correlated noise and uncorrelated noise at a total SNR of 10 to the simulated spectrum, but only include the diagonal of the noise covariance matrix in the likelihood calculation, thereby neglecting the correlation in the noise model. In the ``full matrix'' scenario, the covariance matrix is included in retrievals, which means that the noise is modeled accurately. The corner plot illustrates that parameter uncertainty consistently increases when the noise matrix is included. For example, the 1$\sigma$ confidence interval for Log oxygen abundance increases by a factor of 1.6 when the matrix is included. In addition, all posteriors that do not include the full covariance matrix are significantly biased from truth. This is especially evident by the artificial methane detection when the matrix is not included, which becomes an upper limit when the matrix is included.

These findings replicate the results from \cite{Greco+16}. We thus emphasize the importance of measuring the correlated noise in spectral data and including the associating covariance matrix in retrievals, in order to obtain accurate and un-biased posteriors. We assume the noise will be straightforward to measure in annuli of the image, though this assumption should be reconsidered in future work. The subsequent retrievals presented in this paper will all include the full covariance matrix of the noise.

\subsection{Varying correlation length scale}
\begin{figure*}[t!]
    \centering
    \subcaptionbox{\label{fig:panel1}}{%
        \includegraphics[width=.48\textwidth]{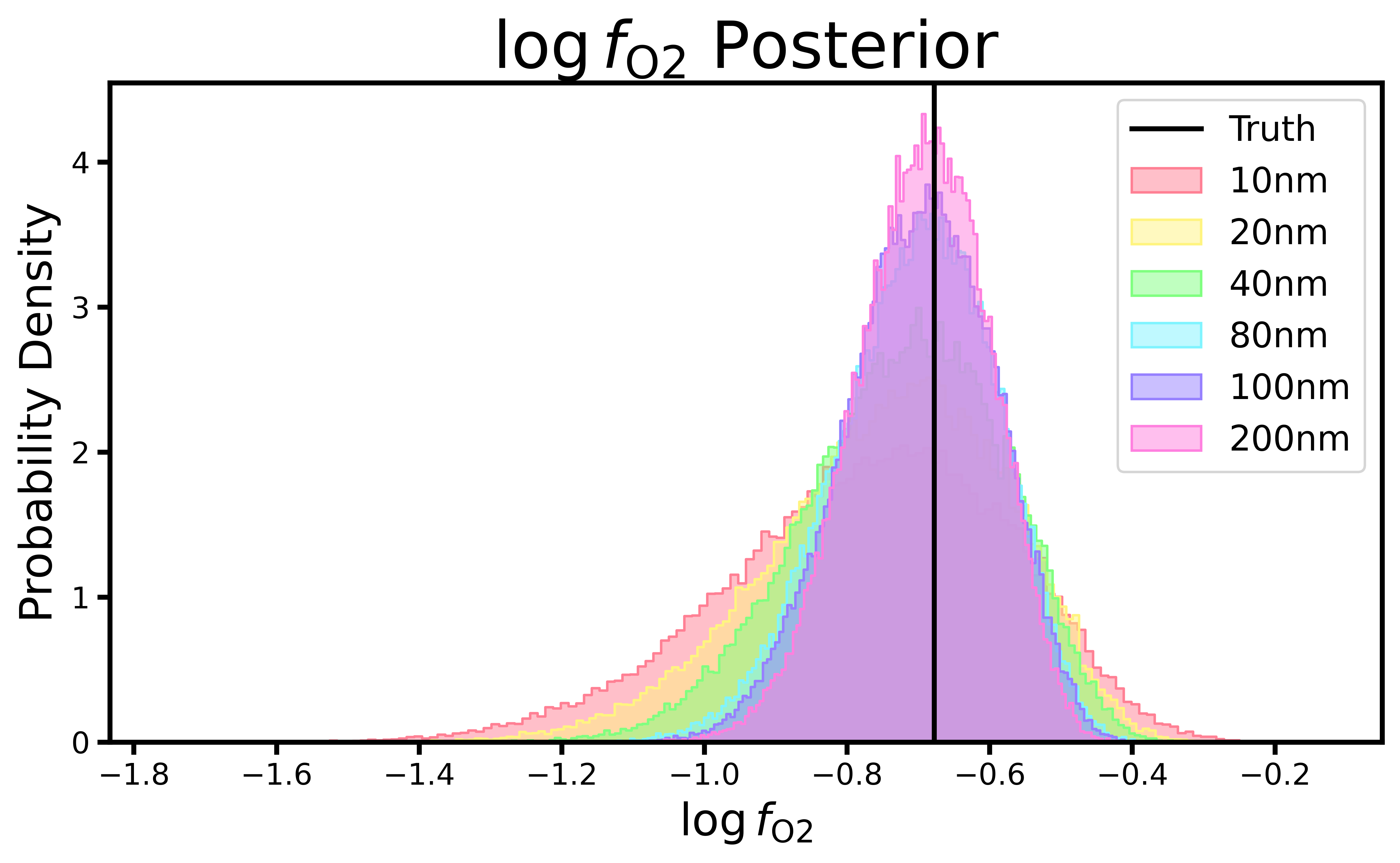}%
    }
    \hfill
    \subcaptionbox{\label{fig:panel1}}{%
        \includegraphics[width=.48\textwidth]{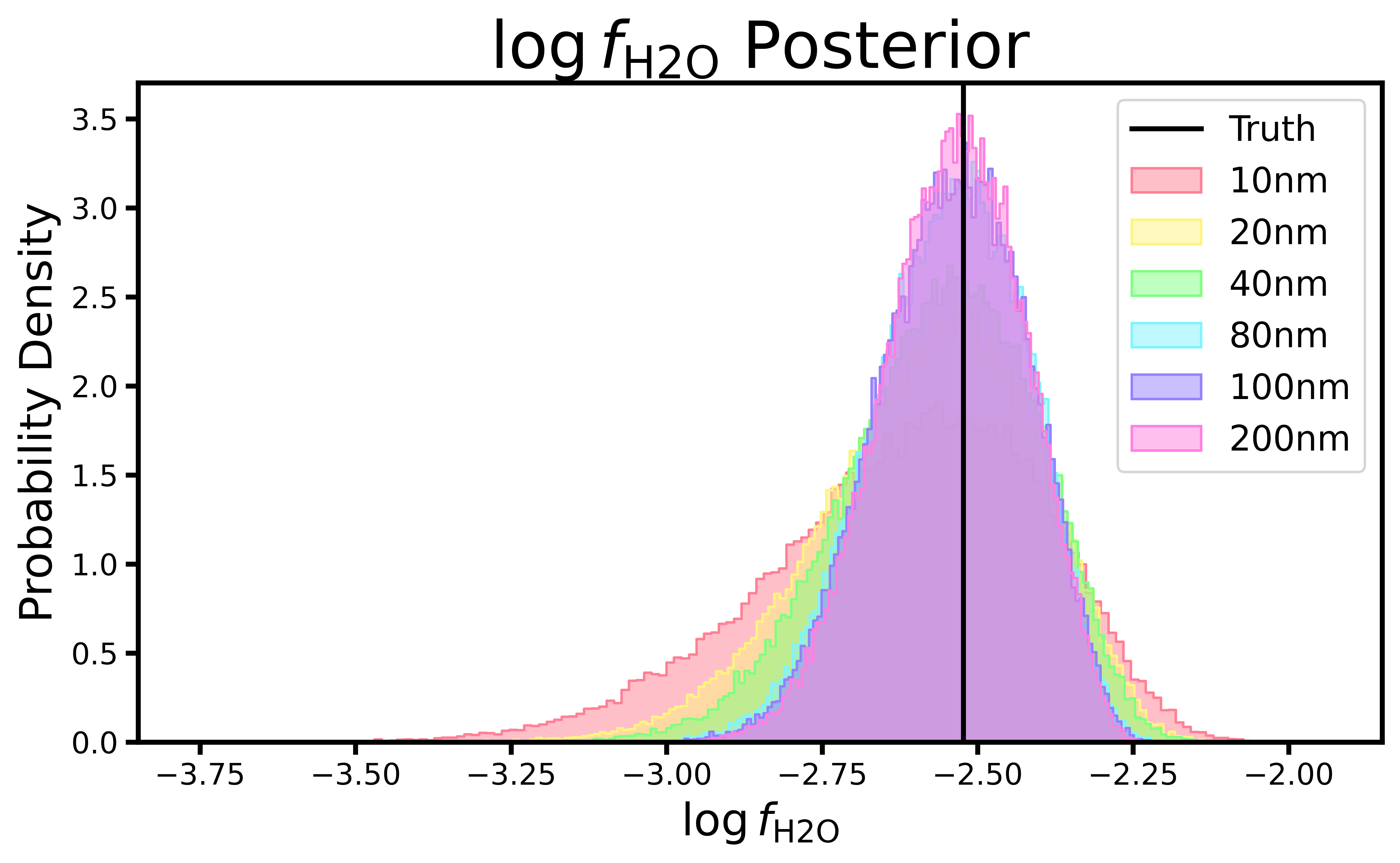}%
    }
    \caption{1-D posteriors for noise correlation length scales of [10, 20, 40, 80, 100, 200] nm are over-plotted. They are normalized such that the area under each histogram is 1. The true values for the abundances in the model are indicated by vertical lines. \textbf{Left:} posteriors of retrieved Log oxygen abundance. \textbf{Right:} posteriors of retrieved Log water abundance.}
    \label{f:length_scale}
\end{figure*}

We present retrieval results across six length scales, $\sigma_c \in$ [10, 20, 40, 80, 100, 200] nm, in order to quantify the effect of length scale on parameter uncertainty. Two parameter posteriors for oxygen and water abundance are shown in Figure \ref{f:length_scale}. For this experiment, we choose not to add correlated noise to the planetary spectra, instead centering the data on truth, in an assumption that the spectrum of the noise is opportunistically flat. The correlated noise is included in the covariance matrix. This method ensures that the posterior median does not fluctuate due to the realization of the noise, and instead remains centered on truth, allowing us to overlay posteriors and visually observe how their uncertainties scale with the covariance matrix.

We observe that the uncertainty scales inversely with length scale. This arises from the mathematical parameterization of the noise (Equation \ref{e:cov_matrix}), such that the magnitude of the covariance is inversely proportional $\sigma_c$, thus validating that our noise model produces the expected results in retrievals.

\subsection{Impact of correlated noise at low resolution}
To test the impacts of correlated noise across a statistical sample of faux observations, we run 20 retrievals, each with a different random realization of the noise. This is to ensure that one particular result is not biased by an extreme random draw of correlated noise. We run these retrievals for two edge cases: a 10 nm and 200 nm length scale, as detailed in Table \ref{t:all_retrievals}. A 10 nm length scale is chosen as a limiting case  because this matches the FWHM of the oxygen absorption line at 760 nm. A 200 nm length scale is chosen because the first-order effect of PSF magnification due to speckles will produce long-scale correlated noise of order 200 nm (Appendix \ref{sec:psfmag}).

\begin{figure*}[t!]
\centering
\subcaptionbox{\label{fig:panel1}}{
    \includegraphics[height=6cm, angle=0]{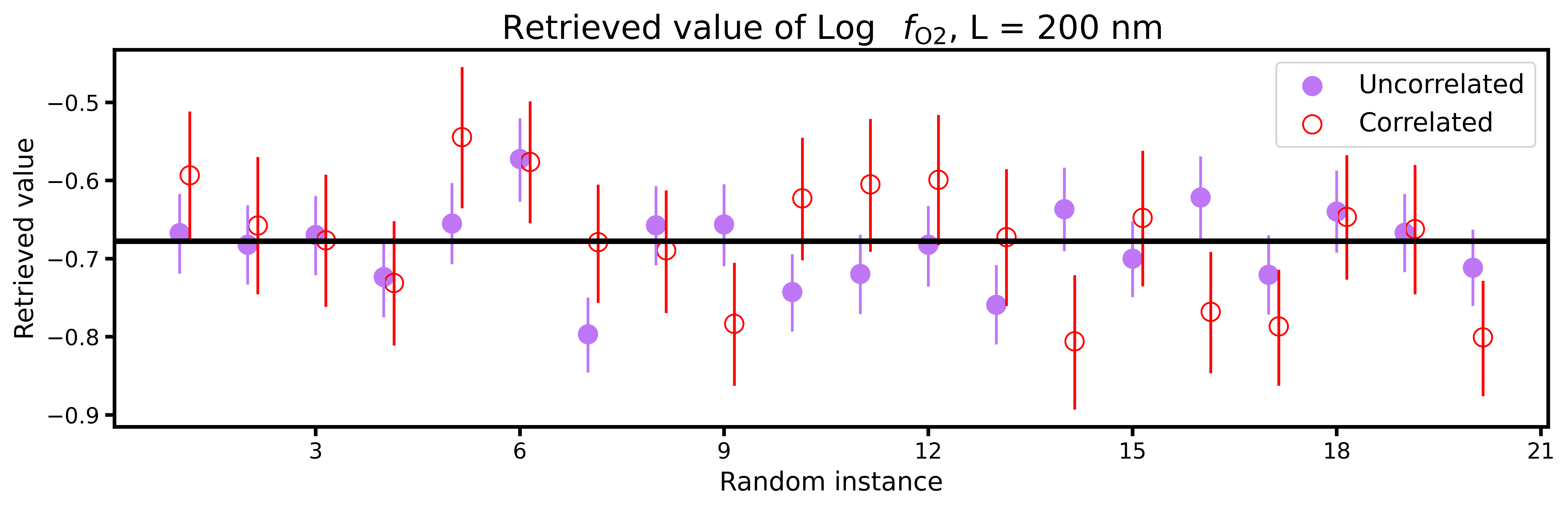}
    }
\hfill
\subcaptionbox{\label{fig:panel1}}{
    \includegraphics[height=6cm, angle=0]{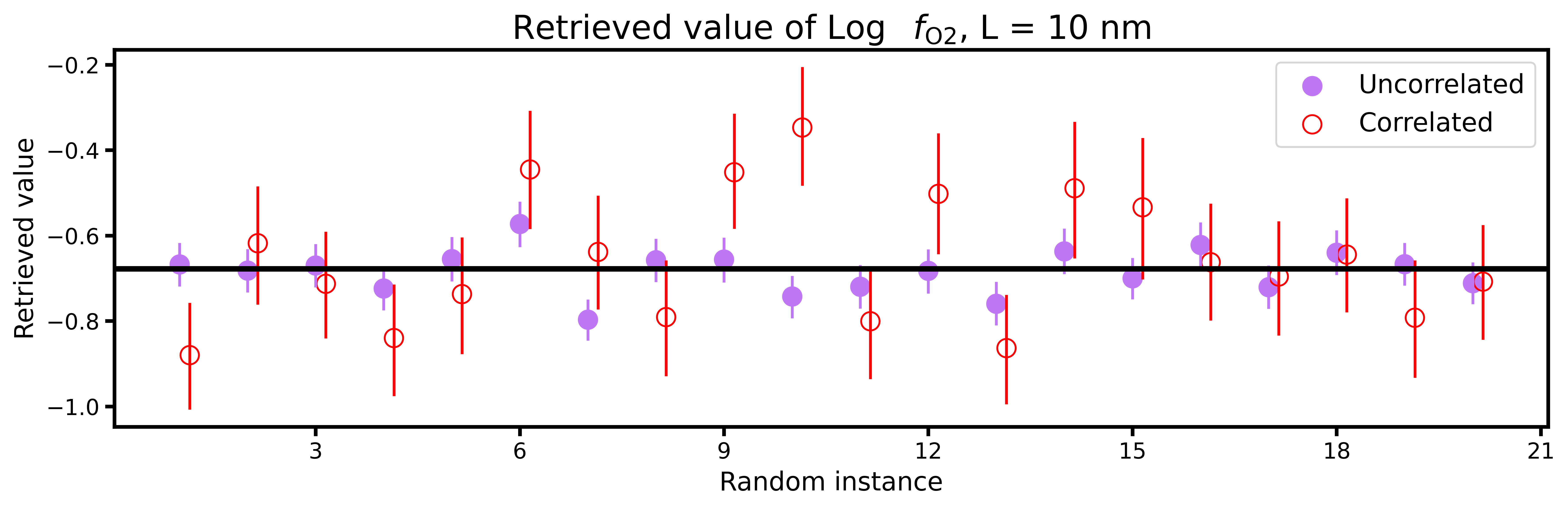}
    }
\caption{Retrieved oxygen abundance across 20 random realizations of correlated noise. Uncorrelated noise (purple, filled) and correlated noise (red, unfilled) retrievals are compared side-by-side with a small horizontal offset. The 1-$\sigma$ confidence interval is shown for each measurement. \textbf{Top:} L = 200 nm. \textbf{Bottom:} L = 10 nm. The black vertical line indicates ground truth.}
  \label{f:bias_l200}
  \vspace{-6pt}
\end{figure*}

Results are presented in Figure \ref{f:bias_l200}. We plot the median retrieved value of oxygen across all noise realizations, along with the 1$\sigma$ confidence interval, both for uncorrelated noise and correlated noise. To directly compare the effect of spectrally correlated noise to the previously-studied scenario of purely uncorrelated noise, we run the uncorrelated and correlated retrievals for a total SNR of 10 in both cases. For the uncorrelated case, the SNR of 10 is purely contained in the uncorrelated noise. For the correlated case, uncorrelated noise is first added to the spectrum and the diagonal of the covariance matrix at an SNR of 28.28, or 20$\sqrt{2}$ (Table \ref{t:all_retrievals}), to simulate a component of uncorrelated photon noise. The remaining and predominant noise contribution comes from correlated noise at an SNR of 10.69. These SNRs sum to a total SNR of 10 (Equation \ref{e:6}), allowing the uncorrelated and correlated cases to be directly compared. This figure illustrates that the purely-uncorrelated case (filled, purple points) consistently yields lower parameter uncertainty than the correlated case (unfilled, red points), signaling that spectrally-correlated noise adversely affects abundance inference.

These results additionally confirm that there is no systematic bias over many realizations of the noise, but an individual sample of residual correlated noise may be significantly offset in any direction. These results additionally confirm that the parameter uncertainty is consistent across realizations of the noise, and that the uncertainty increases in the presence of correlated noise, especially at shorter length scales.

\subsection{Varying spectral resolution, narrow-bandpass}\label{sec:narrowbp}
For the (R,SNR) trade space investigation, we run retrievals across a 20\% band-pass in the visible (650-800 nm), replicating a single coronagraphic observation. We retrieve all parameters listed in Table \ref{t:parameter_inputs}, excluding CH$_4$ (which has no lines in this band-pass). We additionally fit for the planetary mass to determine whether it becomes detectable at higher resolving power. In all cases, mass is unconstrained.
% This is primarily done for computational efficiency, allowing us to explore a large parameter space quickly.

We run retrievals on data of varying spectral resolution, $\mathrm{R} \in $ [140, 280, 1400, 2800], and varying uncorrelated SNR (or equivalently, exposure time), $\mathrm{SNR}_\mathrm{u,R=140} \in $ [10, 20, 40]. The uncorrelated SNR is stated as an R=140-equivalent SNR, because the uncorrelated SNR scales with resolution (Equation \ref{e:6}). These retrievals are additionally run for six correlated noise parameterizations, spanning three correlated SNRs, $\mathrm{SNR}_\mathrm{c} \in $[10, 5, 2] and two length-scales, $\sigma_c \in $ [50, 200] nm.

\begin{figure}
    \begin{minipage}[c]{1.0\linewidth}
        \centering
        \begin{center}
        \includegraphics[height=7.5cm]{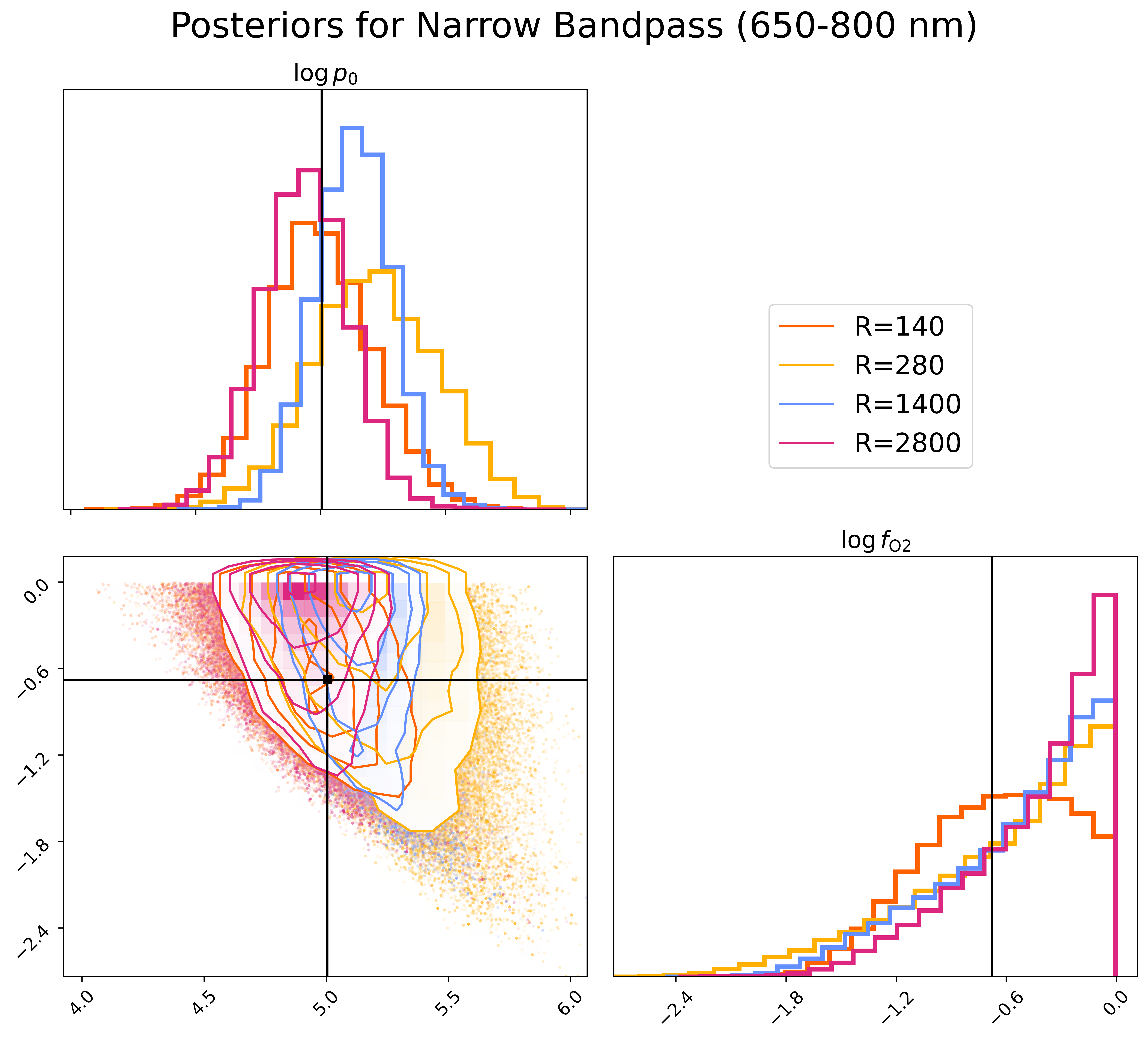}
        \vspace{-5pt}
        \caption{Corner plot displaying the surface pressure-fractional oxygen abundance degeneracy for various spectral resolutions. The noise properties shown here are $\mathrm{SNR}_\mathrm{u,R=140}=40, \mathrm{SNR}_\mathrm{c}=10$. The uncorrelated SNR decreases with spectral resolution such that all cases are at a constant exposure time. In all posteriors, the parameter value is on the x-axis and the probability density is on the y-axis.}
        \label{f:narrow_bandpass_1}
        \vspace{-6pt}
        \end{center}
    \end{minipage}
\end{figure}

Given the large number of retrieval cases run in the narrow-bandpass (72 total), not all confidence intervals are stated in this paper. Across all noise parameterizations and resolving powers, most parameter posteriors
remain relatively unchanged. This is due to the strong degeneracy between surface pressure and molecular abundances, described in Section \ref{sec:observables}. Fig. \ref{f:narrow_bandpass_1} shows a section of a corner plot displaying the degeneracy surface pressure-fractional oxygen abundance for the highest-SNR noise parameterization and all four resolving powers. This corner plot is a representative example of all the narrow-bandpass retrievals, displaying the trend that oxygen is consistently detected but never constrained at all spectral resolutions and SNRs. This result is discussed in Section \ref{sec:narrowbp}.

\subsection{Varying spectral resolution, full bandpass}
\begin{figure*}[h!]
    \begin{minipage}[c]{1.0\linewidth}
  \begin{center}
  \centering
  \includegraphics[width=1.0\linewidth,angle=0]{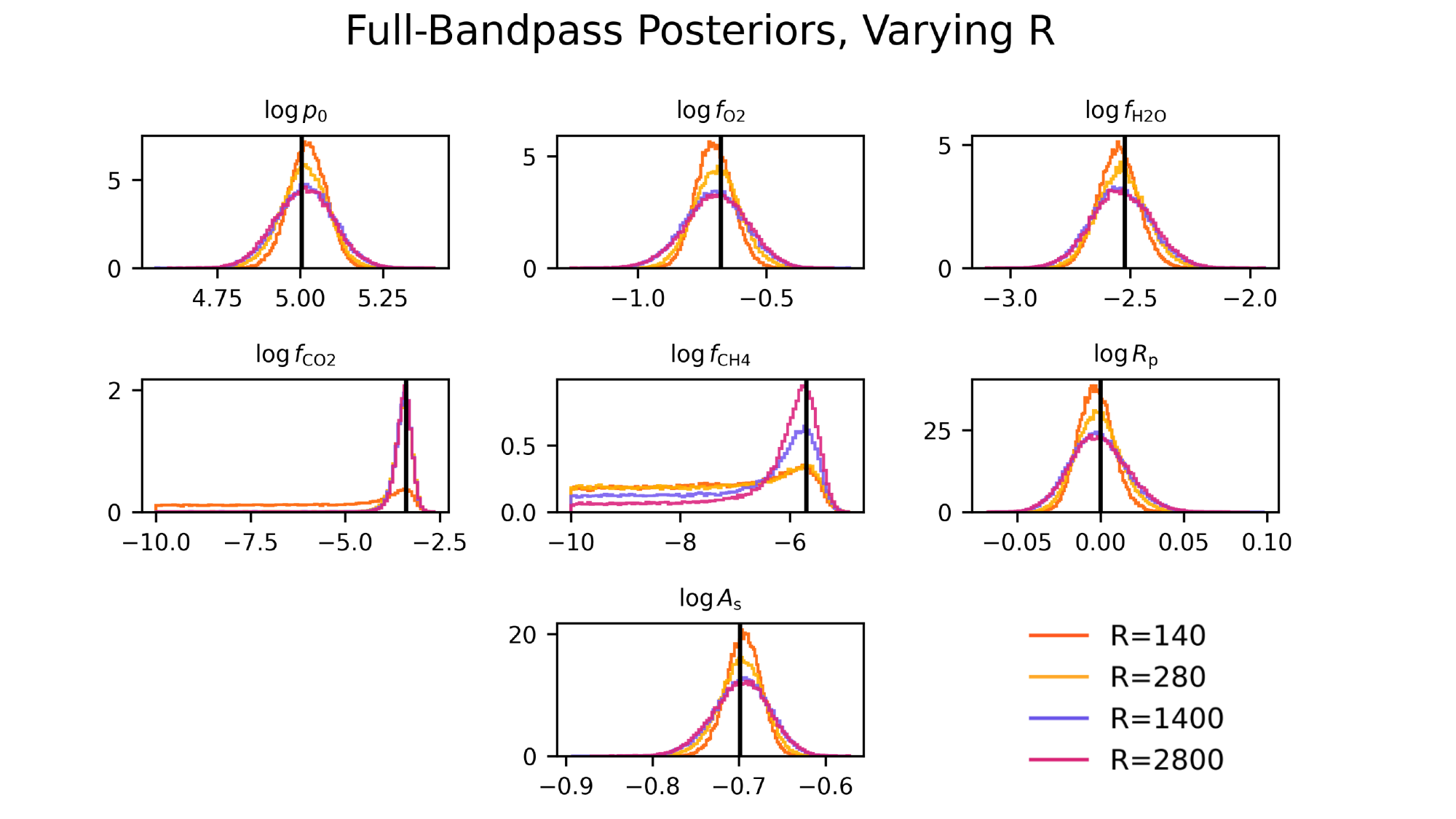}
  \vspace{-5pt}
  \caption{Posteriors of all retrieved parameters as a function of spectral resolution, at $\mathrm{SNR}_\mathrm{u,R=140}=20, \mathrm{SNR}_\mathrm{c}=10$. The uncorrelated SNR decreases with spectral resolution such that all cases are at a constant exposure time. In all posteriors, the parameter value is on the x-axis and the probability density is on the y-axis.}
  \label{f:full_bp_varying_res_all}
  \vspace{-6pt}
  \end{center}
  \end{minipage}
\end{figure*}

\begin{figure*}[t!]
    \centering
    \subcaptionbox{\label{fig:panel1}}{%
        \includegraphics[width=0.49\textwidth]{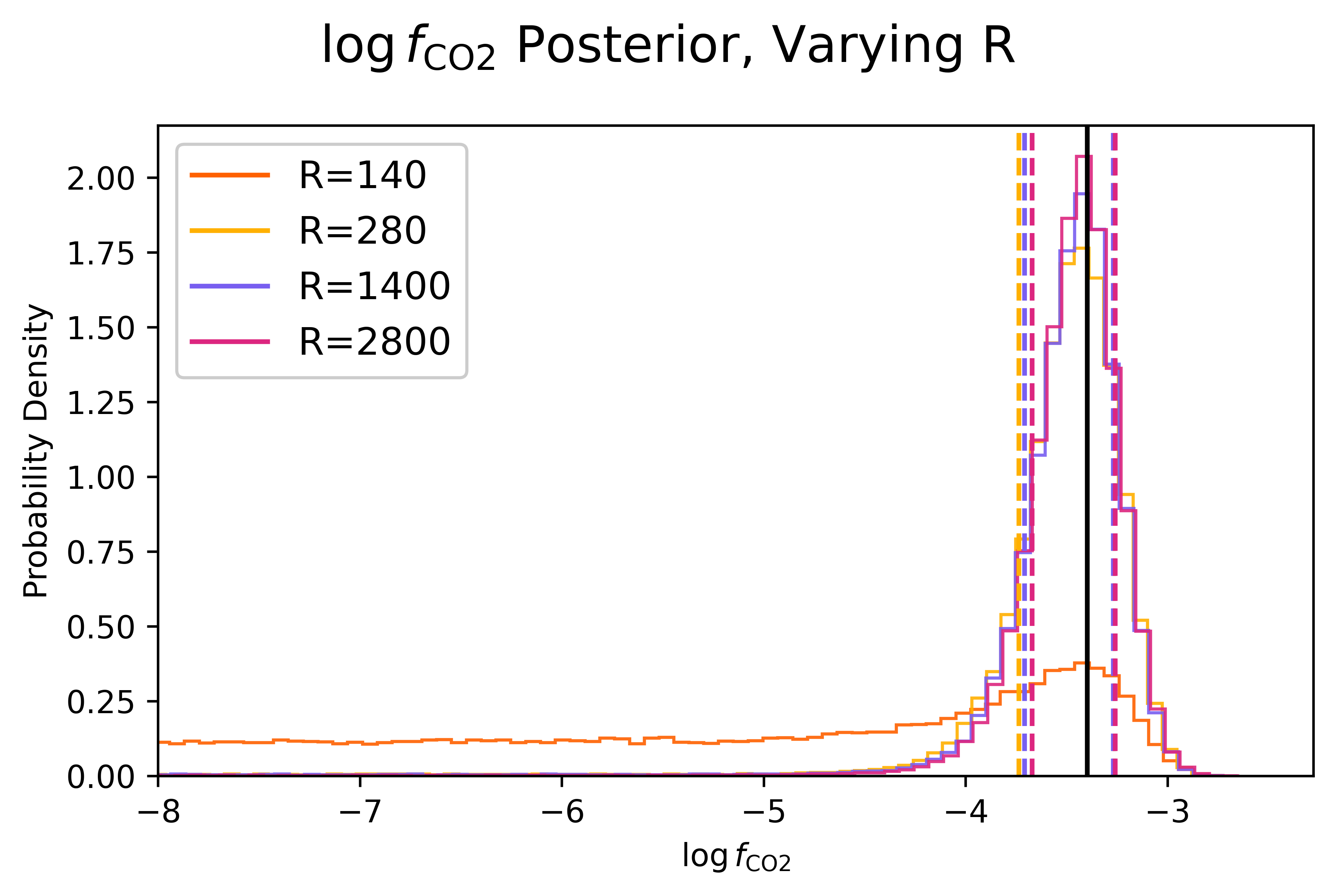}%
    }
    \subcaptionbox{\label{fig:panel3}}{%
        \includegraphics[width=0.49\textwidth]{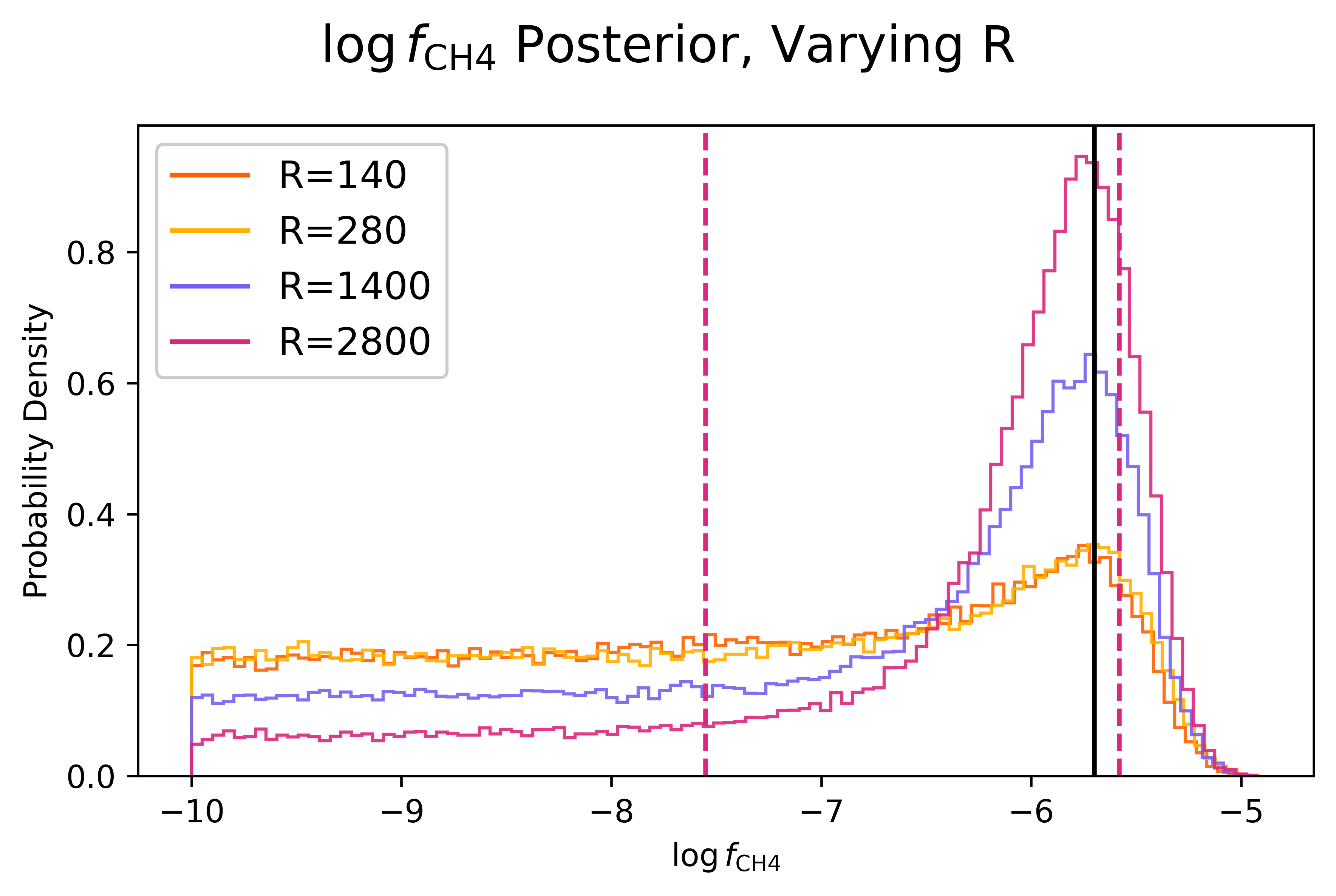}%
    }
    \caption{Posteriors of trace gases in Figure \ref{f:full_bp_varying_res_all} as a function of spectral resolution \textbf{Left:} CO$_2$ posteriors. \textbf{Right:} CH$_4$ posteriors.}
    \label{f:full_bp_varying_res}
\end{figure*}

The final retrievals presented in this study are across the full bandpass, at one slice of the noise parameter space explored in \ref{sec:narrowbp}, and at varying spectral resolution. All posteriors are presented in Figure \ref{f:full_bp_varying_res_all}. For the surface pressure, oxygen, water, radius, and albedo posteriors, the retrieved uncertainty increases with spectral resolution by a non-statistically significant amount (i.e. all parameters remain well-constrained). For the carbon dioxide posterior, R=280 marks the transition from an upper limit to a constraint. For methane, R=2800 marks the transition from an upper limit to a weak detection. The CO$_2$ and CH$_4$ posteriors are magnified in Figure \ref{f:full_bp_varying_res}.

\definecolor{lred}{RGB}{255, 153, 154}
\definecolor{lgreen}{RGB}{205, 255, 154}
\definecolor{lgreen}{RGB}{182, 250, 145}
\definecolor{lyellow}{RGB}{255, 255, 5}
\definecolor{lyellow}{RGB}{247, 232, 20}
\definecolor{gray}{RGB}{211, 211, 211}

\subsection{Summary of results}
We can qualitatively describe the detection significance of an observation based on (a) the magnitude of the retrieved uncertainty and (b) the shape of the resulting posterior. We adopt the formalism in \cite{Feng+18} and \cite{Salvador+Robinson25} for categorizing the detection sensitivity of the model parameters. \textcolor{lgreen}{Constraints} indicate a peaked posterior with a 1-sigma confidence interval less than an order of magnitude; \textcolor{lyellow}{detections (D)} indicate broader posteriors with a 1-sigma confidence interval greater than an order of magnitude, with \textcolor{lyellow}{weak detections (WD)} having additional extended tails towards the prior limits; \textcolor{lred}{upper limit/lower limit (UL/LL)} signifies a flat posterior which excludes a region of the prior above/below a certain value; and \textcolor{lred}{ND} signifies a non-detection.

In Table \ref{t:retrieval_results}, we list the retrieved confidence intervals for all of the aforementioned cases in this work, color-coded by this detection significance formalism. For each case, we only choose to list one slice of the SNR, such that the total SNR is 10. As a result, all cases in the table can be effectively compared.

\begin{deluxetable*}{c|c|c|c|c|c|c|c}
\centering
\tablehead{\textbf{Case} & O${}_2$ & H${}_2$O & CO${}_2$ & CH${}_4$  & Pressure & Radius & Albedo }
\startdata
\hline
Uncorrelated&
{\cellcolor{lgreen}}$-0.67^{+0.05}_{-0.05}$&
{\cellcolor{lgreen}}$-2.59^{+0.06}_{-0.06}$&
\makecell{{\cellcolor{lyellow}}$-5.39^{+1.99}_{-3.11}$ \\ WD}&
{\cellcolor{gray}}N/A&
{\cellcolor{lgreen}}$5.02^{+0.04}_{-0.04}$&
{\cellcolor{lgreen}}$-0.005^{+0.008}_{-0.008}$&
{\cellcolor{lgreen}}$-0.69^{+0.01}_{-0.01}$\\
\hline
\makecell{L=10 nm, \\ no matrix}&
{\cellcolor{lgreen}}$-0.23^{+0.11}_{-0.1}$&
{\cellcolor{lgreen}}$-1.62^{+0.15}_{-0.15}$&
\makecell{{\cellcolor{lred}}$-5.77^{+2.72}_{-2.92}$ \\ UL}&
{\cellcolor{lgreen}}$-4.44^{+0.17}_{-0.22}$&
{\cellcolor{lgreen}}$4.45^{+0.1}_{-0.1}$&
{\cellcolor{lgreen}}$0.11^{+0.02}_{-0.02}$&
{\cellcolor{lgreen}}$-0.91^{+0.05}_{-0.05}$\\
\hline
\makecell{L=10 nm, \\ full matrix}&
{\cellcolor{lgreen}}$-0.78^{+0.18}_{-0.17}$&
{\cellcolor{lgreen}}$-2.72^{+0.23}_{-0.18}$&
\makecell{{\cellcolor{lred}}$-6.12^{+2.57}_{-2.61}$ \\ UL}&
\makecell{{\cellcolor{lred}}$-6.11^{+2.57}_{-2.01}$ \\ UL}&
{\cellcolor{lgreen}}$5.2^{+0.15}_{-0.16}$&
{\cellcolor{lgreen}}$-0.03^{+0.03}_{-0.02}$&
{\cellcolor{lgreen}}$-0.63^{+0.05}_{-0.05}$\\
\hline
\makecell{L=200 nm, \\ full matrix}&
{\cellcolor{lgreen}}$-0.69^{+0.08}_{-0.08}$&
{\cellcolor{lgreen}}$-2.53^{+0.1}_{-0.09}$&
\makecell{{\cellcolor{lred}}$-5.73^{+2.27}_{-2.92}$ \\ UL}&
{\cellcolor{gray}}N/A&
{\cellcolor{lgreen}}$5.03^{+0.06}_{-0.07}$&
{\cellcolor{lgreen}}$-0.005^{+0.01}_{-0.01}$&
{\cellcolor{lgreen}}$-0.69^{+0.02}_{-0.02}$\\
\hline
\makecell{L=200 nm, \\ full matrix, \\ 20\%-Vis \\ bandpass}&
\makecell{{\cellcolor{lred}}$-0.58^{+0.40}_{-0.60}$ \\ LL}& \makecell{{\cellcolor{lyellow}}$-2.65^{+0.43}_{-0.59}$ \\ D}& 
\makecell{{\cellcolor{lred}}$-5.21^{+3.33}_{-3.24}$ \\ ND}&
{\cellcolor{gray}}N/A&
{\cellcolor{lgreen}}$5.10^{+0.36}_{-0.37}$& {\cellcolor{lgreen}}$-0.11^{+0.20}_{-0.14}$& \makecell{{\cellcolor{lyellow}}$-0.46^{+0.29}_{-0.41}$ \\ WD}\\
\hline
\makecell{L=10 nm, \\ full matrix, \\ full-bandpass, \\ R=2800}&
{\cellcolor{lgreen}}$-0.69^{+0.12}_{-0.12}$&
{\cellcolor{lgreen}}$-2.54^{+0.13}_{-0.12}$&
{\cellcolor{lgreen}}$-3.45^{+0.19}_{-0.24}$&
\makecell{{\cellcolor{lyellow}}$-5.97^{+0.39}_{-1.58}$ \\ WD}&
{\cellcolor{lgreen}}$5.01^{+0.09}_{-0.09}$&
{\cellcolor{lgreen}}$-0.002^{+0.02}_{-0.02}$&
{\cellcolor{lgreen}}$-0.7^{+0.03}_{-0.03}$\\
\hline
\enddata
\caption{Parameter confidence intervals for a key selection of the retrieval cases in this paper. \textbf{\textcolor{lgreen}{Constraints}} indicate a peaked posterior with a 1-sigma confidence interval less than an order of magnitude; \textbf{\textcolor{lyellow}{detections (D)}} indicate broader posteriors with a 1-sigma confidence interval greater than an order of magnitude, with \textcolor{lyellow}{weak detections (WD)} having additional extended tails towards the prior limits; \textbf{\textcolor{lred}{upper limit/lower limit (UL/LL)}} signifies a flat posterior which excludes a region of the prior above/below a certain value; and \textbf{\textcolor{lred}{ND}} signifies a non-detection (\citealt{Feng+18}, \citealt{Salvador+Robinson25}).}
\label{t:retrieval_results}
\end{deluxetable*}

\section{Discussion} \label{sec:discussion}

For a mission that aims to find life if it exists in another solar system by constraining biosignature gases abundances and their environmental context, it is crucial that HWO can measure these gas abundances to the highest precision. In this study, we focus on quantifying how correlated noise affects the uncertainty in retrieved parameters, particularly the molecular abundances of oxygen, water, carbon dioxide, and methane.

\subsection{Correlation length scale trade study}
At the fiducial resolving power of R=7/140/70 in the UV/Vis/NIR, we find that the shortest correlation length scale will yield the highest abundance uncertainty (Figure \ref{f:length_scale}). This is mathematically evident from the increased correlation between directly neighboring wavelength channels, as in Equation \ref{e:cov_matrix}. Our findings validate this intuition, but we wish to emphasize the importance for exo-Earth spectroscopy. Oxygen abundance uncertainty for 200 nm correlated noise is, on average, a factor of 1.57 greater than an uncorrelated noise scenario. For 10 nm correlated noise, the average uncertainty is a factor of 2.61 greater than an uncorrelated noise scenario. This inflated uncertainty is problematic for atmospheric characterization, and provides reason to prioritize instrument design choices that mitigate short length-scale chromaticity. 

Retrievals ran across various realizations of the noise, at a shorter and longer length scale, confirm the increased uncertainty at shorter length scales (Figure \ref{f:bias_l200}). These results additionally show that all retrieved values confidence intervals contain truth roughly 67\% of the time, which validates that the correlated noise model is not making parameter posteriors non-Gaussian.

In this work, we assume that each individual noise realization simulated comes from data that has already been co-added across multiple exposures, a process which does not necessarily average out spectral correlations if the noise temporal correlations are long (for example, PSF subtraction with a fainter reference star than the target can result in time-persistent correlated noise). In practice, the noise will additionally have a correlation timescale and will be modeled by a full error budget with various length scales and timescales, such that the covariance matrix is exposure-time dependent.

\subsection{Narrow bandpass trade study}

For the 20\% bandpass retrievals, many of the oxygen posteriors are lower limits rather than Gaussian detections, and are unable to rule out a 100\% O$_2$ atmosphere. This is visible in Figure \ref{f:narrow_bandpass_1} for all resolutions. Surface pressure (or column depth) and fractional oxygen abundance are highly degenerate, and continuum information is required to break this degeneracy. Without information from the continuum over a broad wavelength range, oxygen can be detected but not quantitatively constrained due to the uncertainty on column depth. Equivalently, the retrieval is not able to calibrate the depth of the oxygen line to its fractional abundance. At lower SNRs (thus not seen in this figure), the posterior for pressure/or oxygen is instead bimodal, which we discuss further in Section \ref{sec:bimodal}.

The remaining parameters not depicted in Figure \ref{f:narrow_bandpass_1} have roughly constant posterior forms across the spectral resolution/noise parameterization trade space. H$_2$O is consistently weakly detected at low SNR and detected, not constrained, at high SNR; CO$_2$ is never detected, except an upper limit is achieved at R=1400 and R=2800 across all SNRs; mass is never detected; and planetary radius is consistently constrained or detected with a tail towards higher values, but the limited scattering slope from 650-800 nm does not contain enough information to break the radius-albedo degeneracy. As a result, albedo is often a weak detection, with a peaked posterior but a significant tail towards albedo=0 and/or 1.

We conclude that it is difficult to glean information from retrievals at a narrow bandpass, regardless of spectral resolution. These results extend those in \cite{Ruffio+26_earths}, who find that an increased spectral resolution is always beneficial for oxygen detection and is only hindered past a certain resolution by detector noise, which we do not include in our study. Their work uses the molecular mapping/CCF technique to explore the detectability/non-detectability of a molecular absorption line as the noise and spectral resolution changes. Our work involves a measurement of greater complexity: the full physical state of a planetary atmosphere. We find that our results are limited by parameter degeneracies, particularly that of surface pressure and fractional abundance. An extreme example is illustrated in Figure \ref{f:bimodal}, where two very different atmospheric states produce identical spectra. This is an issue unique to spectral retrievals, and can often be remedied by increasing the bandpass of an observation (e.g. \citealt{Salvador+Robinson25}).

\subsubsection{Bimodal posteriors}
\label{sec:bimodal}
\begin{figure}[h!]
 \begin{minipage}[c]{1.0\linewidth}
  \centering
  \begin{center}
  \includegraphics[height=5cm]{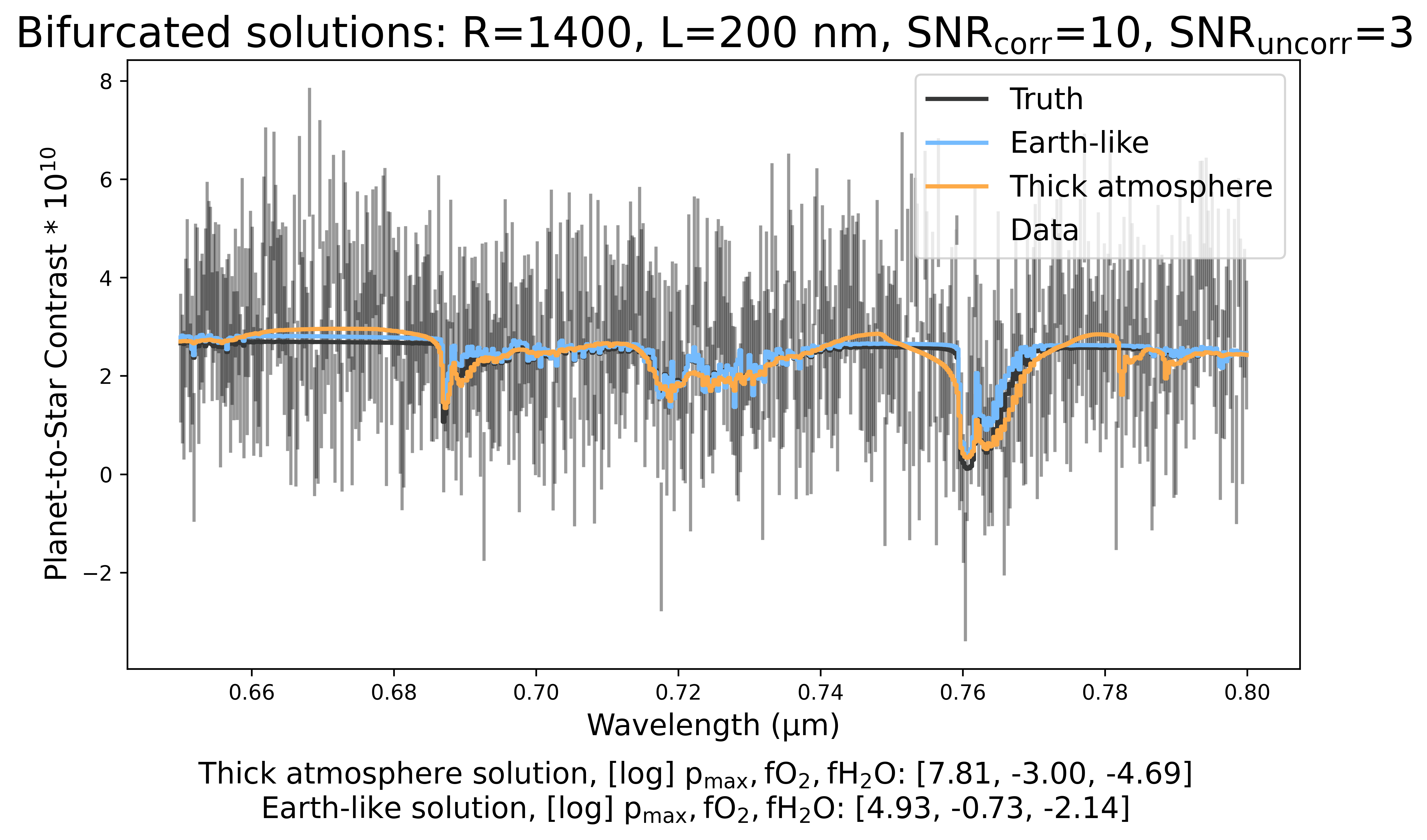}
  \vspace{-5pt}
  \caption{Bifurcated best-fit solutions. The noisy generated data is shown with gray error bars. The true spectrum of Earth is in black, while the Earth-like solution and thick-atmosphere solution are shown in blue and orange, respectively.}
  \label{f:bimodal}
  \vspace{-6pt}
  \end{center}
  \end{minipage}
\end{figure}

In 6 of the 72 narrowband retrievals, the posterior distributions for some combination of pressure, oxygen, and water are bimodal. This indicates that the parameter estimator is unable to distinguish between two atmospheric states. An example of two best-fit models from a bimodal distribution is shown in Figure \ref{f:bimodal}. The two atmospheric states are an Earth-like solution and a high-pressure, thick atmospheric solution, with a surface pressure, water abundance, and oxygen abundance similar to that Venus but with a CO$_2$ fractional abundance $<$ 1\%. Because the predominant atmospheric constituent of this solution is N$_2$, it is implausible that its surface pressure would be so high, indicating an unphysical solution.

For this bifurcated solution, we run the retrieval for an additional 100,000 steps and observe that the walkers stay stationary, indicating the model has converged on both a plausible and implausible solution. This issue arises only in the limited spectral resolution retrievals, rather than the full UV/Visible/NIR band retrievals. Such a result is expected for retrievals lacking blue spectral coverage. The Rayleigh scattering slope is important to break the degeneracy between thick and thin atmospheres; for instance, thick atmospheres will flatten the continuum and imprint absorption features on the continuum, whereas thin atmospheres will exhibit a larger scattering slope and shallower features. This problem is relevant only at the lowest SNRs: here, for all but the R=280 case, bimodal posteriors appear for only the lowest $\mathrm{SNR}_\mathrm{u}$ cases. We can conclude that  bimodal posterior distributions arise from an inability to fit the spectrum at low SNRs, and are more likely to occur for limited bandpasses. 

It is important to note that clouds mute spectral features and thus broaden the parameter space of thick-atmosphere, high-pressure solutions that can fit an Earth-like spectrum in retrievals, making bifurcated solutions more likely. This result is noted in \cite{Robinson+Salvador23} with cloudy retrievals on EPOXI spectra of Earth, showing a bimodality at SNR=10 which  disappears at SNR=20.  Future retrievals including patchy clouds as well as speckle noise will be necessary to explore the full spectral resolution/SNR parameter space for HWO. 

\subsection{Spectral resolution trade study}
Across the full bandpass, we present retrieval results for this higher-correlated SNR slice of the noise ($\mathrm{SNR}_\mathrm{c}=10, \mathrm{SNR}_\mathrm{u,R=140}=20$). Compared to the narrow bandpass results, all parameters are better-constrained with the additional wavelength information (Table \ref{t:retrieval_results}). All posteriors for these retrievals are shown in Figure \ref{f:full_bp_varying_res_all}. In this case, the uncertainty on oxygen/water abundance, radius, and albedo slightly increases with spectral resolution, but not by a statistically significant amount (i.e. all parameters remain well-constrained). From information theory, the constraining power is expected to only improve with increasing spectral resolution. However, retrievals are often strongly complicated by model parameter degeneracies, allowing an improved constraint on one parameter to occasionally yield a worsened constraint on another. Future work involving a larger suite of retrievals across spectral resolution, and at the full bandpass, are ideal to determine the optimal spectral resolution for constraining oxygen and water.

The higher spectral resolution retrievals yield improved constraints on CO$_2$ and CH$_4$, which are shown in greater detail in Figure \ref{f:full_bp_varying_res} along with confidence intervals in the dashed lines. At this SNR, R=280 pushes CO$_2$ from an upper limit to a constraint, and R=2800 pushes CH$_4$ from an upper limit to a weak detection.

Without accounting for correlated noise, a weak detection of CO$_2$ is possible at SNR=10 (Table \ref{t:all_retrievals}) However, when accounting for correlated noise across the full bandpass, or across a a narrow bandpass in the visible, $\mathrm{CO_2}$ is consistently an upper limit (Table \ref{t:all_retrievals}). As its molecular band structure becomes more prominent at higher spectral resolutions, $\mathrm{CO_2}$ quickly becomes detectable at R $\geq$ 280. A CO$_2$ constraint would provide evidence for a carbon cycle on an exo-Earth, contextualizing any detections of oxygen and water.

Additionally, a CH$_4$ detection, which becomes possible at R $\geq$ 2800 at this SNR, can be interpreted as a sign of disequilibrium chemistry. This provides a robust biosignature when detected alongside with oxygen, given the short lifetime of methane in oxygenated atmospheres \citep{KrissansenTotton+18}. A methane detection can thus rule out false positive O$_2$ scenarios and attribute an O$_2$ abundance to biological forcing. These results suggests that HWO equipped with a moderate resolution NIR spectrograph would have the capabilities to characterize interacting chemical networks unique to organic systems, instead of relying on single molecules subject to potential abiotic false positives.

In order to detect spectral features in the high correlated noise regime, it is common practice to high-pass filter moderate-resolution spectra to remove the continuum prior to molecular detection via template matching. This technique has proven beneficial for giant planet direct spectroscopy with JWST/NIRSpec at R~2,700 \citep{Ruffio+24}. The downside of this strategy is that it removes the continuum, which is required to measure radius and albedo (a detection of Rayleigh scattering in the UV/blue visible spectral band helps break the degeneracy between radius and surface albedo (Figure \ref{f:exoearths})). Thus, if the correlated noise contribution is too large, additional photometric or low-resolution spectroscopic points in the visible band will prove complementary to moderate resolution spectra in the near-infrared. 

\subsection{Correlated noise properties}\label{sec:correlated_noise}

This work is an initial attempt at addressing the issue of correlated noise for HWO across a broad parameter space. These results illustrate that, at minimum, it is necessary to include a covariance matrix in retrievals for HWO trade studies in order to obtain realistic parameter uncertainties. Preliminary studies can include a simplistic Gaussian functional form of the noise, as done here. However, near the inner working angle (IWA), as will be relevant to HWO, the spatial distribution of speckles grows sparse \citep{Mawet+14}. As such, this prescription of correlated noise is a simplification. Speckles will not necessarily appear at high density across a spectrum, but there may rather be one or two speckles that overlap with the planet at specific wavelengths. A future extension of this work should include injecting local correlated noise (i.e. an individual spectrally-Gaussian speckle, rather than a full realization of a Gaussian Process) at individual wavelengths where they might appear, such as at the edges of a coronagraphic bandpass. Additionally, more realistic prescriptions of the noise can be measured from end-to-end simulations for the different telescope architectures and coronagraphs for HWO, which would provide more nuanced results. 

Another important factor worth future consideration is that the ultimate correlated noise source will be from residual speckles after post-processing, which necessarily alters the properties of the noise measured from optical simulations. A more realistic noise model would take the form of a spectrally and temporally correlated noise error budget, with terms of varying time scale, length scale, and amplitude which arise from different sources (telescope vibrations, optical effects, NCPAs, PSF subtraction, etc). Future studies should explore the effects of multiple terms of correlated noise, e.g. a shorter and longer length scale simultaneously, as well as the plausibility to fit for these terms in MCMC. However, these results alone illustrate that, even if short length-scale noise is not the dominant component of the noise error budget, it is the most impactful to abundance measurements, even at a higher $\mathrm{SNR_{c}}$ = 10.

One physical source of short-length scale chromatic noise is due to the Talbot Effect. Pure phase aberrations away from the pupil plane in an optical system transform to amplitude aberrations periodically, across a length scale known as a Talbot length \citep{Marois+06}. A single deformable mirror (DM) can correct phase aberrations, but not amplitude aberrations. The primary DM in an optical system is conjugate with the primary mirror,  thus correcting any of its polishing imperfections. A second DM can be placed conjugate with another particularly problematic optic in order to correct for this effect, but a DM will not exist to correct for every optic in the instrument. As such, there will remain some residual amplitude aberrations in the wavefront of light. The Talbot length ($\tau$) is calculated as follows:

\begin{equation}
    \tau = \frac{2 k^2}{\lambda},
\end{equation}

where k is the spatial frequency of the aberration and $\lambda$ is the wavelength of light. Across the wavelength range of HWO and for $k \approx$ mm, typical Talbot lengths range from 1 to 10 m, meaning there are two regimes: one where a minuscule amount of wavefront error transforms from phase to amplitude ($\tau \approx 10 $ m, lower spatial frequencies), and one where the aberration oscillates completely from phase and amplitude ($\tau \approx 1$ m, higher spatial frequencies), the latter regime resulting in a source of spectrally correlated noise. We leave a full derivation of the correlation length scale due to the Talbot effect to future work. However, end-to-end simulations of the Roman CGI coronagraph reveal chromatic noise due to the Talbot Effect at length scales $\approx$ 10 nm, which depend on both the wavelength and the spatial location in the image (see \citealt{Krist+23}, Figure 80). Thus, we believe the length scale of 10 nm we considered here to be physically-grounded, though full bandpass end-to-end simulations of the HWO coronagraph will be required for a more accurate trade study.

\section{Conclusions} \label{sec:conclusions}
Our primary takeaways can be summarized as follows: 
\begin{itemize}
\item Retrieval studies exploring the parameter space of HWO instrument design should account for realistic abundance uncertainties by factoring in spectrally-correlated noise. This can be done as a first pass by including a Gaussian noise covariance matrix in the log-likelihood calculation.
\item The temporal evolution of speckles can produce spectrally correlated noise. This temporal evolution drives a challenging telescope stability requirement. Therefore, analyzing the effect of correlated noise on molecular abundance measurement is critical for the design of the telescope architecture and deformable mirrors.
\item As the limiting noise source for the HWO at low spectral resolutions, correlated residual speckle noise in exoplanet spectra increases uncertainty on atmospheric abundances. Its effect is especially pronounced at the shortest correlation length scales when the correlated noise is most similar to the molecular spectral features. A longer 200 nm and shorter 10 nm correlation length scale yield a 57\% and 161\% higher average uncertainty on Log oxygen abundance, respectively, than uncorrelated noise. 
\item A moderate spectral resolution significantly enhances carbon dioxide and methane detectability in an Earth analog atmosphere, allowing for a carbon dioxide constraint (R $\geq$ 280) and a methane weak detection (R $\geq$ 2800) in the presence of spectrally correlated noise. This preliminary spectral resolution trade study provides reason to consider a moderate-resolution spectrograph for exo-Earth characterization with HWO, especially in the near-infrared, in order to mitigate the effects of residual speckles.

\end{itemize}

\section{Acknowledgments}
This work and a paper by Ji Wang, et al. on a similar topic were independently submitted to AAS Journals. Both papers were independent investigations into the impact of correlated noise on spectral retrievals for HWO. We thank the authors for their cordiality. Material presented in this work is partially supported by the National Aeronautics and Space Administration under Grants/Contracts/Agreements No. 80NSSC25K7300 (B.D., J.-B.R.) issued through the Astrophysics Division of the Science Mission Directorate. T.D.R. gratefully acknowledges support from NASA's Habitable Worlds Program (No.~80NSSC20K0226), NASA's Exoplanet Research Program (No.~80NSSC25K7149), and the Nexus for Exoplanet System Science Virtual Planetary Laboratory (No.~80NSSC23K1398).

\software{
 \texttt{astropy} \citep{astropy:2013, astropy:2018, astropy:2022}, \texttt{corner} \citep{corner}, \texttt{emcee} \citep{emcee}, \texttt{matplotlib} \citep{matplotlib}, \texttt{numpy} \citep{numpy}, \texttt{rfast} \citep{Robinson+Salvador23}, \texttt{scipy} \citep{scipy}}

\appendix

\section{PSF magnification}\label{sec:psfmag}

The correlation length scale, visualized in Figure \ref{f:speckle}, measures how much a speckle passing over the planet couples neighboring spectral points.  Across multi-wavelength images, a planet is physically stationary, while a speckle moves outward with wavelength. Here, we derive how correlation length scale depends on the physical separation of the planet:  

\begin{equation}
    \rho_\textrm{planet} = \text{n}_\text{planet}\frac{\lambda_0}{\text{D}}
\end{equation}

This term is dependent on a reference wavelength, $\lambda_0$, as well as the dimensionless term, $\mathrm{n}_\mathrm{planet}$, which denotes the planet's position in diffraction units. Because speckles both grow and move outward with wavelength, 

We first define the intensity of a single speckle as a function of spatial separation and wavelength: 

\begin{equation}
    \text{I}_\text{speckle}(\rho_\text{speckle},\mathrm{\lambda}) = \text{A}^2 \ \text{exp} \left[- 
\frac{(\rho_\text{speckle} - \rho_\text{planet})^2}{2\sigma_\text{speckle}^2}\right]
\end{equation}

\begin{equation}
  \rho_\text{speckle} = \text{n}_\text{speckle}  \frac{\ \lambda}{\text{D} } 
\end{equation}

For the speckle's intensity in (A2) to follow a diffraction-limited PSF, the speckle's width is given as:

\begin{equation}
    \sigma_\text{speckle} = 1.22\frac{1}{\sqrt{8 \ \text{ln} 2} } \frac{\lambda}{ \text{D}}
\end{equation}

Rewriting (A2),

\begin{equation}
    \text{I}_\text{speckle}(\rho_\text{speckle},\mathrm{\lambda}) = \text{A}^2 \ \text{exp} \left[- \frac{1}{2}
\left(\frac{ \text{n}_\text{speckle}  \frac{\ \lambda}{\text{D} }  - \rho_\text{planet}}{\sigma_\text{speckle}}\right)^2\right]
\end{equation}

Setting $\text{n}_\text{speckle} = \text{n}_\text{planet}$,

\begin{equation}
    \text{I}_\text{speckle}(\mathrm{\lambda}) = \text{A}^2 \ \text{exp} \left[- \frac{1}{2}
\left(\frac{ \lambda - \rho_\text{planet} \frac{D}{\text{n}_\text{speckle}}}{\frac{D}{\text{n}_\text{speckle}}\sigma_\text{speckle}}\right)^2\right]
\end{equation}

Given that the planet has a finite size, $\frac{\lambda_0}{\text{D}}$, the problem of generating a spectrum from a 2D image of overlapping speckle \& planet data can be framed as a a convolution between the speckle's flux intensity and a Gaussian kernel with FWHM $\lambda / \text{D}$ (Equation 7). When $\lambda$ is a free variable, these functions transform from spatially-dependent to wavelength-dependent, resulting in speckle intensity across the planet as a function of wavelength (Equation 8).

\begin{equation}
    \text{F}(\lambda) = \beta^2 \ \text{exp} \left[- \frac{1}{2}\left(\frac{\lambda - \lambda^\prime}{\text{D}\frac{1.22}{\sqrt{8 \ \text{ln} 2} } \frac{\lambda}{ \text{D}}
}\right)^2\right]
\end{equation}

\begin{equation}
    \text{S}(\lambda) = \text{I}_\text{speckle}(\lambda) * \text{F}(\lambda)
\end{equation}

We are interested in the width of $\text{S}(\lambda)$. The convolution of two Gaussians with widths $\sigma_1, \sigma_2$ is a Gaussian with width:

\begin{equation}
    \sigma_\mathrm{convolved}^2 = \frac{\sigma_1^2 \ \sigma_2^2}{\sigma_1^2 + \sigma_2^2} 
\end{equation}

Finally, we can write $\Lambda_\mathrm{corr} = \sigma_\mathrm{convolved}$, using $\sigma_1$ and $\sigma_2$ from Equations 6 \& 7:

% \begin{equation}
%     \Lambda_\text{corr}^2 =\frac{(1.22 \lambda_0)^2}{8 \ \ln 2 (1 + \text{n}_\text{speckle}^2)} 
% \end{equation}
% (update)
\begin{equation}
    \Lambda_{\mathrm{corr}}= \frac{1.22}{2 \sqrt{2 \ln 2} } \frac{\lambda^2}{\mathrm{D} * \Theta_\mathrm{planet}}
\end{equation}

At smaller radial separations, a speckle will pass across the planet for a longer duration in wavelength space, resulting in a longer correlation length. (Thus, Figure \ref{f:speckle} is a simplifying illustration, as it does not depict the changing length scale with separation).

\bibliography{sample7}{}
\bibliographystyle{aasjournalv7}

\end{document}